# Linear and Nonlinear Rheology and Structural Relaxation in Dense Glassy and Jammed Soft Repulsive Microgel Suspensions

Ashesh Ghosh[1,+], Gaurav Chaudhary[2,+], Jin Gu Kang[3], Paul V. Braun[1-5], Randy H. Ewoldt[2,4,5*] and Kenneth S. Schweizer[1,3-6*]

[1] Department of Chemistry, University of Illinois at Urbana-Champaign, Urbana, IL, 61801

[2] Department of Mechanical Science and Engineering, University of Illinois at Urbana-Champaign, Urbana, IL, 61801

[3] Department of Materials Science and Engineering, University of Illinois at Urbana-Champaign, Urbana, IL, 61801

[4] Frederick Seitz Materials Research Laboratory, University of Illinois at Urbana-Champaign, Urbana, IL, 61801, USA

[5] Beckman Institute for Advanced Science and Technology, University of Illinois at Urbana-Champaign, Urbana, IL, 61801, USA

[6] Department of Chemical & Biomolecular Engineering, University of Illinois at Urbana-Champaign, Urbana, IL, 61801, USA

*kschweiz@illinois.edu

*ewoldt@illinois.edu

+ these authors contributed equally to this work

## Abstract

We present an integrated experimental and quantitative theoretical study of the mechanics of self-crosslinked, neutral, repulsive pNIPAM microgel suspensions over a very wide range of concentrations ($c$) that span the fluid, glassy and putative "soft jammed" regimes. In the glassy regime we measure a linear elastic dynamic shear modulus over 3 decades which follows an apparent power law concentration dependence $G' \sim c^{5.64}$, a variation that appears distinct from prior studies of crosslinked ionic microgel suspensions. At very high concentrations there is a sharp crossover to a nearly linear growth of the modulus. To theoretically understand these observations, we formulate an approach to address all three regimes within a single conceptual Brownian dynamics framework. A minimalist single particle description is constructed that allows microgel size to vary with concentration due to steric de-swelling effects. Using a Hertzian repulsion interparticle potential and a suite of statistical mechanical theories, quantitative predictions under quiescent conditions of microgel collective structure, dynamic localization length, elastic modulus, and the structural relaxation time are made. Based on a constant inter-particle repulsion strength parameter which is determined by requiring the theory to reproduce the linear elastic shear modulus over the entire concentration regime, we demonstrate good agreement between theory and experiment. Testable predictions are then made. We also measured nonlinear rheological properties with a focus on the yield stress and strain. A theoretical analysis with no adjustable parameters predicts how quiescent structural relaxation time changes under deformation, and how the yield stress and strain change as a function of concentration. Reasonable agreement with our observations is obtained. To the best of our knowledge, this is the first attempt to quantitatively understand structure, quiescent relaxation and shear elasticity, and nonlinear yielding of dense microgel suspensions using microscopic force based theoretical methods that include activated hopping processes. We expect our approach will be useful for other soft polymeric particle suspensions in the core-shell family.

## I. Introduction

Colloidal suspensions have been a major area of interest in the soft matter community for decades. Much fundamental research has been done with model hard-sphere colloids, with or without small polymer depletants, which have elucidated the understanding of physical phenomenon such as crystallization, phase separation, glassy dynamics, and nonlinear rheology [1-3]. Other widely studied systems are dense suspensions of soft colloids [4,5]. However, they bring additional complexities since the particles are deformable with a fluctuating internal polymeric microstructure, which can lead to their size and even shape becoming a function of thermodynamic state (volume fraction, temperature, ionic strength) and deformation. Most microgels are charged and can be created with diverse chemistry, which introduces concentration-dependent complexities associated with osmotic decompression, the poorly known internal density profile (often core-shell), and variable single particle mechanical stiffness. Hence, the effective interaction potential between microgel particles is a complex issue, consistent with a lack of universal signatures in their rheology [6,7]. Moreover, microgels can exist as dense Brownian suspensions that can form kinetic glasses or gels, or at ultra-high concentration as paste-like materials characterized by literal contacts between deformed particles. If the latter exist, the system is typically viewed as in a "soft jammed" regime. However, whether the physics in this regime is entirely akin to granular materials where large scale motion requires the application of external mechanical energy is not well understood, and the answer may depend on system and thermodynamic state.

In this paper, we perform a coordinated experimental and theoretical study of the dynamics and rheology of soft, thermoresponsive poly(N-isopropylacrylamide) (pNIPAM) based microgel suspensions under conditions where they are swollen in a good solvent and repel. There have been extensive prior studies of similar systems [5, 8-11], albeit mainly in the soft jamming regime with ionic microgels which are chemically crosslinked and can osmotically de-swell with changing concentration [5, 8-9]. Such microgel pastes are generally viewed as effectively athermal or granular.

Our study has several not very common features: (a) there is no chemical crosslinking via added molecules of the microgel particles, (b) the microgels are strictly uncharged, and (c) experiments are performed over an exceptionally wide range of concentration that spans the low viscosity fluid, glassy Brownian, and soft jammed regimes. These aspects distinguish our experimental system from most others, and isolates particle compression as *solely* due to many body steric effects. We will show that points (a) and (b) lead to mechanical behavior with features significantly different than prior studies. Point (c) is also a strong focus of this work where in the first two concentration regimes there are no literal inter-particle "contacts" and the mechanical response is influenced by Brownian caging processes driven by thermal fluctuations and external stress [12-13]. The possibility that the ultra-dense regime is not granular-like is also explored.

The remainder of this article is structured as follows. In section II we describe the experimental materials and methods. Our key experimental results for the linear and nonlinear rheology are presented in section III. Section IV presents the basics of our theoretical modeling of single microgel structure, and the equilibrium and dynamical statistical mechanical tools we employ to make predictions for collective packing structure, linear elasticity, structural relaxation, and aspects of nonlinear rheology. Quantification of microgel effective volume fraction is discussed in section V, and predictions made for the linear dynamic shear modulus and packing structure, with the former compared with our measurements. Theoretical results for the equilibrium structural relaxation time, its variation with deformation, and yielding properties are presented in Section VI, with some comparison to experiment. The paper concludes with a discussion in Section VII. Additional experimental characterization and rheological results are presented in the Supplementary Information (SI).

## II. Materials and Methods

### A. Microgel synthesis and characterization

Neutral self-crosslinked pNIPAM microgels were synthesized under a 'crosslinker free' condition following the protocol described in literature [14] with modifications (see Supplementary section 1 for details). Free-radical polymerization of NIPAM in water was initiated using potassium persulfate in the absence of added cross-linker. This leads to the formation of stable nanospheres instead of linear chains if the solution is incubated at temperatures well above the lower critical solution temperature (LCST) of PNIPAM (~32℃). The formation of gel nanospheres is attributed to self-crosslinking by chain transfer reaction during and after polymerization [15]. A stock solution of $c = 9\,wt\%$ was then diluted with deionized water to achieve the desired concentration of the uncharged microgel suspension.

The particle radius was determined by dynamic light scattering (DLS) (Zetasizer Nano ZS, Malvern) and a Helium-Neon gas laser emitting at $632.8\,nm$ on a very dilute suspension ($0.04\,wt\%$) with a beam diameter of $0.63\,mm$ (See Supplementary Figure S1). The present work focuses on the lower temperature regime where microgels are swollen and interact via repulsive forces. In dilute solution, the microgel particles have a mean diameter of $2R = 551 \pm 71\,nm$ at 10℃.

### B. Rheological Characterization

Rheological experiments are performed over a very wide range of microgel concentration from $0.03\,wt\%$ to $9\,wt\%$. Viscoelasticity was probed using a rotational rheometer (model

Discovery Hybrid 3, TA instruments and model MCR702 from Anton Paar) with plate-plate geometry. These are both torque-controlled instruments (a.k.a. combined-motor-transducer type). A $600$ grit, adhesive-back sand paper (Norton Abrasives) was adhered to the contact surfaces to suppress wall slip. The plate diameter was varied depending on the sample concentration to obtain a measurable response higher than the minimum torque resolution. A $60\ mm$ plate was used for dilute samples $0.03 - 0.25\ wt\%$, $40\ mm$ plate for $(0.25 - 1.5)\ wt\%$, $20\ mm$ for $(0.5 - 4.5)\ wt\%$, and $8\ mm$ for $(4.5 - 9)\ wt\%$ samples. The typical gap in all experiments was between $(550 - 750)\ \mu m$, far larger than the particle size, thus eliminating confinement effects. A solvent trap, with a wet-tissue adhered to its interior, was used to minimize solvent evaporation during the measurements. The temperature of the bottom plate was controlled using a Peltier-system. To suppress sample aging effects and erase any history, all samples were rejuvenated by shearing at $50\ s^{-1}$ for $60s$ and then allowed to relax for $12\ min$ before taking measurements [5].

Two types of rheological characterization were performed: oscillatory shear and steady shear. To probe the linear response, frequency sweeps were performed from $\omega = (0.03 - 100)\ rad/s$ at a strain amplitude of $1\%$ at 10°C. To probe the nonlinear response, strain sweeps of amplitude $\gamma_0 = (0.1 - 300)\%$ at a fixed frequency of $1\ rad/s$ were performed. In the steady shear experiments, shear rates were typically varied from $(300 - 0.01)\ 1/s$ while waiting for the system to reach an apparent steady state as deduced by $< 5\%$ variation in torque over a period of $30\ s$.

## III. Experimental Results

### A. Linear Rheology

Figure 1A shows the frequency-dependent linear storage, $G'$, and loss , $G''$, moduli as a function of frequency. One sees a nearly frequency independent $G'$, with a smaller $G''$ that also is weakly frequency-dependent. Hence, $G'' < G'$ for all concentrations above $0.4\, wt\%$ and the response is predominantly solid-like with the structural or flow relaxation time obeying $\tau_\alpha > \omega_{low}^{-1} \approx 100\ s$. No crossover between $G'$ and $G''$ was observed in the range of frequencies probed, indicating the microgels do not show significant diffusion or structural relaxation on the probing time scales.

At higher frequencies, the commonly observed frequency dependence of $G'' \sim \omega^{1/2}$ for a loosely and randomly packed emulsion is very roughly observed for the $0.75\, wt\%$ and $1\, wt\%$ samples [16]. However, there are systematic deviations -- power laws are often not well developed, and apparent scaling exponents, if force fit, can be larger or smaller than 0.5, and tend to decrease as concentration grows. For concentrations below $1\, wt\%$, the inertia of the measuring system influences the torque measurements significantly and makes it difficult to observe any reliable signatures [17] for high frequency measurements.

The linear storage modulus at a fixed frequency of $\omega = 1\ rad/s$ and a strain amplitude of $\gamma_0 = 1\%$ is shown in Figure 1B. It monotonically grows with increasing concentration (as also found at slightly higher temperatures, see Supplementary Figure S3). Three distinct regimes of behavior are observed. For concentrations below $c = 0.4\, wt\%$, no measurable elastic modulus is detected above the minimum torque limit of the instrument. This seems consistent with measurements of the high shear rate viscosity (Supplementary Figure S2), where an excellent agreement with the Einstein prediction of the dilute intrinsic viscosity is observed in the concentration range $(0.03 - 0.35)\ wt\%$, beyond which the viscosity strongly grows. Since the microgels are neutral, the latter is presumably due to repulsive inter-microgel forces and transient caging in the suspension. Such a fundamental change in the concentration range $(0.4 - 0.5)\ wt\%$ is consistent with a dynamic crossover to a regime where there is little particle motion on the experimentally probed time scales [11,18]. In hard sphere glasses the characteristic modulus scale is set by the thermal energy per particle volume [4,13], $G \sim k_B T/(2R)^3$, where $k_B$ is Boltzmann's constant, $T$ is temperature, and $R$ is the particle radius, which for our system is $G' = 0.024\ Pa$ for $2R = 550nm$. This estimate is fairly close to when we first observe a solid-like response: $G' = 0.04\ Pa$ and $G' = 0.14\ Pa$ for $0.4\ wt\%$ and $0.5\ wt\%$ concentrations, respectively.

In the intermediate concentration range, defined as $(0.4 - 1.25)\ wt\%$, the elastic modulus shows a dramatic dependence on microgel concentration. A variance weighted fit of all data yields $G' \sim c^{5.64 \pm 0.28}$, but it seems clear the effective exponent weakly decreases with concentration. Similar observations have been made in literature [5,11], but the apparent power law exponent in Fig. 1B is generally very different for previous work using pNIPAM based suspensions (see Supplementary Figure S4 for comparison). For example, Menut et al. [5] observed power law exponents of 4.4, 6.1 and 14, respectively, for three p(NIPAAm-co-AAc) ionic microgel suspensions of increasing single particle stiffness as synthesized by precipitation polymerization with varying cross-linker concentration. Pellet and Cloitre [11] observed a power-law exponent of 9.1 for a suspension of polyelectrolyte microgels synthesized by emulsion polymerization. Given the narrow range of data in the "glassy regime" of that study, such a high apparent exponent may simply indicate exponential growth.

In the highest concentration range of our experiments, defined as $(1.5 - 9)\,wt\%$, the elastic response again qualitatively changes. The modulus now grows weakly in a nearly linear manner with concentration (variance weighted fit, $G' \sim c^{1.17 \pm 0.07}$). How to interpret this solely from mechanical data is neither obvious nor unique. We can envision three possibilities. (1) It could indicate a transition to what is usually called a "soft jammed" state where microgels are in literal contact, particles may deform and form facets, and elastic energy is stored in a granular manner. This scenario predicts $G' \propto (\phi - \phi_{jam})$ [11], which to be consistent with our data seems to require the effective volume fraction grows linearly with microgel concentration (which is a priori unclear). (2) Discrete microgel particles could somehow effectively "fuse" in the practical sense that the suspension behaves as a connected macroscopic network of flexible "elastically active chains or strands". From the classical theory of rubber elasticity, this scenario implies elasticity is fundamentally of single strand (conformational) entropic origin, with $G' \sim \rho_x kT$ where $\rho_x$ is the polymer concentration divided by the number of monomer units in each polymer strand, $N_x$ [5,19]. A comparison between our experimental data and the rubber elasticity model [20] is given by the red line (variance weighted fit parameter, $N_x = 435$) in Figure 1B. (3) A third scenario is the change in concentration dependence of $G'$ reflects a crossover from sterically-induced weak compression of core-shell microgels to a regime where the microgels isotropically shrink in a manner that keeps its effective volume fraction fixed. This scenario retains the discrete picture of microgel particles, does not invoke facets or literal interparticle contacts, and posits an interparticle collective origin of stress storage. It will theoretically be developed in section IV, and shown to also lead to a linear growth of $G'$ with microgel concentration. While we cannot completely rule out there might be components of scenario (1) or (2) that contribute to the observed linear growth of elastic modulus of our system, in this article we take a minimalist approach of exploring a Brownian glassy suspension scenario for the *entire* concentration regime without invoking athermal soft jamming.

Supplementary Figure S4 shows elastic modulus data from other labs for different types of microgels, all of which are ionic. Clearly, one sees that at fixed concentration in $wt\%$, different microgel samples display a wide variety of modulus levels and sensitivity to concentration. This emphasizes that our present self-crosslinked neutral microgel system with different chemistry does display a distinct elastic response. Figure S4 also emphasizes the far larger range of concentration probed in our study (factor ~25) versus prior studies (typically factor of 3-10). However, these prior studies all observe, to varying degrees, a stronger growth of $G'$ at lower concentration followed by a much weaker growth at very high microgel concentrations.

**B. Nonlinear Rheology**

Our nonlinear oscillatory shear measurements are shown in Figure 2. Only the first-harmonic responses are plotted, representing the average storage and loss of mechanical energy, here indicated as $G_1'$ and $G_1''$, respectively. The response at all concentrations is similar. At low strains, the response is in the linear regime, with roughly a constant value of $G_1'$ and $G_1''$ and $G_1' > G_1''$. At large strains, the response becomes nonlinear with $G_1'$ showing a monotonic decrease while $G_1''$ exhibits a maximum. An increasing $G_1''$ indicates more dissipation presumably due to deformation-induced microgel motion which can be qualitatively viewed as a stress driven solid-to-fluid like transition or yielding [5,13]. One measure of the latter is the strain at which $G_1' = G_1''$, which occurs at rather high strain values of $\sim 25 - 50\%$ with systematic variation with concentration difficult to discern. More precise definitions and analysis of yielding will be given in section VI.

Figure 3A shows the steady state flow curve of the microgel suspensions. Below $c = 0.4\,wt\%$, the response resembles a shear thinning fluid at high shear rates. At higher concentrations, $c > 0.4\,wt\%$, the stress-strain rate response resembles that of a yield-stress fluid, although for most samples there is no rigorous low shear plateau and the degree to which the data is flat does not vary systematically with concentration. Such a response can be adequately captured by the empirical Herschel-Bulkley (HB) model given by: $\sigma(\dot{\gamma}) = \sigma_y^{\mathrm{HB}} + K(\dot{\gamma})^n$, where $\sigma_y^{\mathrm{HB}}$ is the apparent yield strength, $n$ is the flow index, and $K(\dot{\gamma})^n$ describes the shear-thinning behavior at high shear rates for $n < 1$ [13]. The parameter $K$ has dimensions that depend on $n$ and does not represent a physical quantity. However, we can instead use a modified form of the HB model [21],

$$\sigma(\dot{\gamma}) = \sigma_y^{HB}\left(1 + \left(\frac{\dot{\gamma}}{\dot{\gamma}_c}\right)^n\right) \tag{1}$$

where the characteristic shear rate, $\dot{\gamma}_c = \left(\frac{\sigma_y^{\mathrm{HB}}}{K}\right)^{1/n}$, is associated with a crossover from rate-independent plastic flow to rate-dependent viscous flow. Equation (1) is used to fit the experimental data which directly yields the parameter $\dot{\gamma}_c$.

The HB fits to the data (assuming constant error weighting) and corresponding fit parameters ($\sigma_y^{HB}, \dot{\gamma}_c\,, n$) are shown in Figure 4. Similar to the observations made earlier for the linear elastic modulus, we find a strong concentration dependence of $\sigma_y^{HB} \sim c^{4.5}$ in the intermediate concentration regime, which is however clearly weaker than that of the $G'$ data in Fig.1B. We will refer to such behavior as indicating the "glassy regime". At higher concentrations the yield stress grows roughly linearly with concentration, which for descriptive purposes we refer to as the "soft jamming" regime. The flow index, $n$, decreases monotonically with the concentration in the glassy regime, $n \sim c^{-0.48}$, followed by a nearly constant value of $0.41$ in the soft jamming regime. The lower inset of Fig.4 shows that the characteristic shear rate $\dot{\gamma}_c$ is roughly constant in the glassy regime and follows a power law relation, $\dot{\gamma}_c \sim c^{-2.5}$, in the soft jammed regime. As true of the linear elastic modulus, Figure 4 shows that the yielding properties of our microgel suspensions follow quite different trends from previous studies [11] of different ionic microgel systems. Specifically, the yield stress in the soft jamming regime displays a stronger concentration dependence ($\sim c^2$), the exponent $n$ values are generally larger, and $\dot{\gamma}_c$ increases with the concentration in the glassy regime until appearing to plateau in the soft jamming regime.

## IV. Theoretical Approach: Microgel Model, Packing, Elasticity, Dynamics, and Rheology

### A. Overview and Modeling of Single Microgel Structure in the Condensed Phase

Much theoretical progress has been made in recent years by many workers [22-25] for understanding the slow dynamics and rheology of simple colloidal particles which can be treated as soft or hard spheres that interact via a central pair potential, $V(r)$ [6]. If $V(r)$ is known, then one can use a litany of statistical mechanical methods to analyze their collective structure, equilibrium dynamics under Brownian conditions, and nonlinear rheology. The approach Schweizer and co-workers have developed and widely applied in prior work [26] is used here and proceeds in a series manner as follows. (1) Construct a single particle model and $V(r)$. (2) Use liquid state integral equation methods to predict the intermolecular pair correlation function, $g(r)$, and its Fourier space collective structure factor, $S(k)$. (3) Use (1) and (2) to construct predictive dynamical theories of thermally activated equilibrium structural relaxation dynamics and mechanical properties, and (4) combine (1)-(3) to construct a theory for the effect of deformation on non-equilibrium dynamics and mechanics.

The daunting difficulty to quantitatively carryout such a program for microgels is that they are soft fluctuating polymeric particles with many internal degrees of freedom. Quantitative knowledge of their internal structure in dense suspensions, as a function of thermodynamic state variables (concentration, temperature), is scarce. This renders an a priori theoretical analysis at the monomer level very difficult or impossible. It has led to almost all theoretical and simulation studies adopting a coarse-grained center-of-mass (CM) level description of the polymer microgel which interacts via a pair decomposable isotropic soft repulsive potential where the influence of all internal degrees of freedom are effectively pre-averaged. This corresponds to $V(r)$ becoming a free energy or potential-of-mean force (PMF) quantity. But an a priori quantitative theoretical construction of such a PMF for real chemical systems is extremely difficult since it requires the following information. (i) How a global measure of mean size (radius, $R$) of a single microgel changes as a function of concentration and temperature, i.e. what is $R(c,T)$ ? (ii) What is the functional form of $V(r)$ and how does it change with thermodynamic state? (iii) Even for a simple $V(r)$ such as the Herztian contact model (see below), the single particle modulus is variable, depending on chemistry, preparation method, and crosslink density, and is a priori unknown. (iv) How does the experimental concentration variable (weight percent) map to an effective volume fraction as a function of concentration and temperature, i.e. $\phi_{eff}(c,T)$?

The inability to a priori answer the above questions forces one to adopt models constrained by incomplete knowledge. Physical ideas must be invoked, and parameters introduced, with the goal of retaining some predictive power. Here we outline our approach, which is summarized in Figure 5.

Soft microgels are generally globally compact and compressible objects that are swollen in a good solvent but have a (dense) core - (more dilute/hairy) corona structure [24,27]. We take a microgel to be, on average, a spherical soft object. Its internal density $\rho(r)$ decreases continuously in a non-universal manner upon transitioning from its center to edge, ultimately becoming effectively zero at $r = R_{eff}$. In the dilute low concentration regime the microgel size is fixed and one can define a volume fraction $\phi = \frac{4\pi}{3}\rho R^3$ which grows linearly with concentration. As suggested by experiments of Schurtenberger et. al. [7,28], in an intermediate concentration regime of $c_1 < c < c_2$ (per the notation of Fig.5) the microgels begin to de-swell due to steric repulsions between particles, in a manner that experiments suggest is initially weak. Crudely, experimental data in the latter regime can be modeled as a power law, $R \sim c^{-1/x}$ where $x > 3$, implying an effective volume fraction that scales as $\phi \sim c^{(1-\frac{3}{x})}$. Motivated by the experimental data of Figure 5A of ref. [28], we estimate $x = 1/6$. Beyond a "high enough" $c \geq c_2$, one expects the more fuzzy "corona" of the microgel is largely squeezed out, leaving a dense core which further decreases in size as concentration grows due to isotropic compression in the sense that $x = 1/3$, as again suggested in ref [28]. This leads to $\phi \sim c^0$ where the linear growth of microgel particle number density ($\rho$) with concentration is perfectly compensated by their shrinking size. Ultimately, beyond an even higher concentration $c_3$, the internal concentration of microgels presumably saturates at a maximum value akin to a collapsed molten globule with radius $R_{collapsed}$.

Quantitative knowledge of such a complex, continuous, and material-specific variation of microgel size with concentration is unknown for our system. Thus, we adopt the model of Fig. 5 which has 3 crossover concentrations, one exponent parameter "$x$", and 3 characteristic sizes. The crossover concentrations are determined using our elastic modulus data and theory as explained in detail in section V. Here we summarize the model adopted there.

We assume that the lowest concentration regime extends up to $c = 0.4\ wt\%$ and the microgel size is constant and the same as in the $c \to 0$ dilute limit as determined from our DLS measurements, $2R = 2R_0 = 551$ nm. A second regime is defined starting at $c_1$ (0.4 wt%) and ending at $c_2 = 1.5$ wt% (onset of "soft jamming" behavior of $G'$). Here we assume the microgels begin to weakly contract and employ $x = 1/6$ as suggested in ref. [28]. This implies at $c_2$ the microgel diameter is $2R = 442\ nm$. Beyond $c_2$ a third regime is entered and we adopt the 1/3 exponent to describe microgel shrinkage. This implies at the highest concentration we study (9 wt%) one has $2R \approx 244\ nm$. Interestingly, as Fig. S1 shows, this is roughly the size of dilute microgels at high temperature beyond the LCST where they undergo an enthalpy-driven collapse. Although a collapsed microgel driven by poor solvent conditions need not be exactly the same size as what can be attained via interparticle steric repulsion, it is not unreasonable they could be similar. Hence, in terms of the scenario of Fig. 5 we deduce as a rough approximation $c_3 \sim 9\ wt\%$, and our present measurements do not probe the ultra-high concentration fourth regime which may be impossible to explore in practice.

We employ a suite of older and recently developed theoretical tools to model our system. The rest of this section provides a brief summary without derivation of these methods. All details can be found in original papers, and for consistency we employ the same notation developed in these prior theoretical works. Our present work is the first time the new activated dynamics (ECNLE) theory in equilibrium and under deformation has been employed to study soft colloids.

### B. Center-of-Mass Hertzian Repulsion Model

The vast majority of modeling studies of soft microgels has employed the repulsive Hertzian contact or harmonic interaction model. We adopt the former which is given by [22,29],

$$\beta V(r) = \begin{cases} \frac{4E}{15}\left(1-\frac{r}{d}\right)^{\frac{5}{2}} & if\ r < d = 2R_{eff} \\ 0 & if\ r \geq d \end{cases} \tag{2}$$

where $\beta = (k_B T)^{-1}$ is the inverse thermal energy, $r$ is the interparticle separation, and $d$ is the particle diameter. The front factor ${}^{4E}/_{15}$ is the inverse dimensionless temperature that controls the elastic stiffness of a particle and hence repulsion strength. $E$ is a priori unknown for our system, and $d \approx 2R$ where $R \approx R_g$ of the core-corona particle. From its mechanics derivation, $E$ is related to the sphere diameter $d$, Young's modulus $Y$, and Poisson ratio $\nu$, as:

$$E = \frac{Yd^3}{2k_B T(1-\nu^2)}. \tag{3}$$

Depending on the magnitude of the dimensionless temperature, the Hertzian potential can describe ultra-soft microgels ($E \leq 10^3$), intermediate soft microgels ($10^3 \leq E \leq 10^6$), and effective hard spheres ($E \geq 10^6$). The literal hard sphere limit is smoothly obtained as $E \to \infty$. Very recent simulations of soft microgel suspensions that explicitly considered the polymeric internal degrees of freedom found the Hertzian pair potential to work fairly well [30].

### C. Equilibrium Packing Structure

We use the standard Ornstein-Zernike (OZ) integral equation [31, 32] approach to compute the inter-particle pair structure. The OZ equation relates the non-random part of the interparticle pair correlation function, $h(r) = g(r) - 1$ (where $g(r)$ is the pair correlation or radial distribution function), and the direct correlation function, $c(r)$ via [31, 32],

$$h(r) = c(r) + \rho \int c\left(\left|\vec{r} - \vec{r'}\right|\right) h(r') d\vec{r}' \tag{4}$$

where $\rho$ is the particle number density. Collective density fluctuations are controlled by the static structure factor which in Fourier space is

$$S(k) = 1 + \rho h(k) = \frac{1}{1-\rho C(k)}. \tag{5}$$

Numerical solution of the OZ equation requires a closure approximation that relates $c(r)$, $g(r), V(r)$, and thermodynamic state (density, temperature). For soft colloids the hypernetted chain closure (HNC) relation works well and is given by,

$$c(r) = -\beta V(r) - ln\big(g(r)\big) + h(r) \quad (6)$$

**D. Dynamic Localization and Elasticity: Naive Mode Coupling Theory**

The starting point for describing the dynamics of a tagged particle in a liquid is the Generalized Langevin Equation (GLE) for its position and velocity [31,32],

$$m\frac{d\vec{V}(t)}{dt} + \zeta_s\vec{V}(t) = -\frac{\beta}{3}\int_0^\infty d\tau\,\langle\vec{f}_\alpha(t).\vec{f}_\alpha(t-\tau)\rangle\; + \delta\overrightarrow{f_\alpha}(t)\; + \xi(t) \quad (7)$$

where $\zeta_s$ is a short time friction constant, $\vec{f}_\alpha(t)$ is the force on a tagged particle due to the surrounding particles, and $\delta\overrightarrow{f_\alpha}(t)$ and $\xi(t)$ represent the random white noise (Gaussian) force associated with the short time process. The naive ideal Mode-Coupling Theory (NMCT) of single particle dynamics calculates the force-force time correlation function or memory function by quantifying dynamical constraints at the pair structural level as [26]:

$$K(t) = \langle\vec{f}_\alpha(0).\vec{f}_\alpha(t)\rangle = \frac{\beta^{-2}}{3}\int_0^\infty \frac{\overrightarrow{dk}}{(2\pi)^3}\rho\left|\vec{M}_{NMCT}(k)\right|^2 S(k)\Gamma_s(k,t)\Gamma_c(k,t) \quad (8)$$

where $\vec{M}_{NMCT}(k) = kC(k)\hat{k}$ is the wave vector resolved effective force on a tagged particle, and the "dynamic propagators" $\Gamma_s(k,t), \Gamma_c(k,t)$ are the $t = 0$ normalized single and collective dynamic structure factors (decay to zero in a fluid phase, non-zero for solids). At long times, localized states can exist and the Gaussian Debye-Waller factors are non-zero, $\Gamma_s(k, t \to \infty) = e^{-k^2 r_L^2/6}$, where $r_L$ is the dynamic localization length associated with a kinetically arrested state. The collective propagator is accounted for in a de Gennes narrowing manner as [33],

$$\Gamma_c(k, t \to \infty) \equiv \Gamma_s\left(\frac{k}{\sqrt{S(k)}}, \infty\right) = e^{-k^2 r_L^2 / 6S(k)}. \tag{9}$$

A self-consistent equation in the long time limit for the particle displacement can be derived and is given by: $\beta\langle \vec{f}_\alpha(0).\vec{f}_\alpha(t \to \infty)\rangle^{r_L^2}/_2 = \frac{3k_BT}{2}$ . From this, the ideal NMCT self-consistent localization equation is [34]

$$\frac{1}{r_L^2} = \frac{\rho}{18\pi^2}\int_0^\infty dk\, k^4 C(k)^2 S(k) e^{-\frac{k^2 r_L^2}{6}(1+S^{-1}(k))}. \tag{10}$$

One can also compute the elastic shear modulus associated with such an ideal glass state. The calculation is relevant in practice if the product of the frequency of the measurement and structural relaxation time obeys $\omega\tau_\alpha \gg 1$. A standard statistical mechanical formula for the dynamic elastic shear modulus, based on projecting microscopic stress onto a bilinear product of the collective density fields followed by factorization of multi-point correlations to the 2-point level, is [31]:

$$G' = \frac{k_BT}{60\pi^2}\int_0^\infty dk\, \left[k^2 \frac{d}{dk} ln\big(S(k)\big)\right]^2 e^{-k^2 r_L^2 / 3S(k)} \approx a\phi \frac{k_BT}{d\, r_L^2} = a(\rho k_BT)\left(\frac{d}{r_L}\right)^2 \tag{11}$$

where "a" is a numerical prefactor. The final approximate "microrheology-like" relation can be analytically derived for hard spheres and works well for Hertzian spheres [12]. Tighter dynamic localization (smaller $r_L$) leads to higher mechanical stiffness.

We comment that one might interpret Eq. (11) as suggesting an apparent equivalence of the basic mathematical form of the “microrheology-like” relation to that of classic rubber elasticity, $G' \sim \rho_x kT$. But, there is no conceptual correspondence since $\rho$ is the number of microgels per unit volume in Eq.(11) and not the crosslink number density as in rubber elasticity. Moreover, the localization length is an emergent dynamic quantity associated with kinetic trapping of particles due to interparticle forces and is a strong function of the thermodynamic state variables.

Most fundamentally, the basis of Eq. (11) is the spatial correlation in a (transiently in practice) kinetically arrested state of collective *inter*particle microscopic stress defined by particle positions and interparticle forces, not the intra-strand entropic stress per rubber elasticity.

### E. Quiescent Activated Structural Relaxation

To go beyond ideal MCT to treat thermally activated events that lead to slow structural relaxation, the nonlinear Langevin equation (NLE) theory has been developed. It is based on the scalar displacement of a tagged particle, $r(t)$, as the central dynamic variable. In the overdamped limit, the stochastic NLE for a particle trajectory is [33,34]

$$\zeta_s \frac{dr}{dt} = -\frac{\partial F_{dyn}(r)}{\partial r} + \xi(t) \tag{12}$$

where $\xi(t)$ is a Gaussian white noise and the key quantity is the *dynamic* free energy, $F_{dyn}$. The gradient of the latter determines the instantaneous force on a moving tagged particle due to its near neighbors and is given by [34]

$$\beta F_{dyn}(r) = \frac{3}{2} ln\left(\frac{3d^2}{2r^2}\right) - \frac{\rho}{2\pi^2}\int_0^\infty dk\ \frac{k^2 C(k)^2 S(k)}{(1+S^{-1}(k))} e^{-\frac{k^2 r_L^2}{6}(1+S^{-1}(k))}. \tag{13}$$

The first contribution is an ideal entropy like term that favors the delocalized fluid state, and the second interaction free energy like term favors dynamic localization. The dynamic free energy is constructed to recover NMCT per $\left.\frac{\partial F_{dyn}(r)}{\partial r}\right|_{r=r_L} = 0$. At and above a critical volume fraction $\phi > \phi_c$ ($\approx$ 0.43 for hard spheres [34]) a barrier in $F_{dyn}(r)$ emerges at $r = r_B$ of height $F_B$ with a corresponding transient localization length $r_L$; see Figure 6 for an example. The liquid structural relaxation time is estimated from the Kramers mean barrier hopping time as [34]

$$\frac{\tau_\alpha}{\tau_s} = 1 + \frac{2}{d^2}\int_{r_L}^{r_B} dx\, e^{\beta F_{dyn}(x)} \int_0^x dy\, e^{-\beta F_{dyn}(y)} \approx 1 + \frac{2\pi}{\sqrt{K_0 K_B}} e^{\beta F_B} \tag{14}$$

where $\tau_s$ is a short time process relaxation time and $K_0$ and $K_B$ are positive local curvatures of free energy at $r_L$ and $r_B$, respectively. The approximate relation in Eq. (13) holds when $\beta F_B \gtrsim 1-2$. The short time scale is [35]:

$$\tau_s = g(d)\frac{d^2}{D_{SE}}\left[1+\frac{1}{36\pi\phi}\int_0^\infty dQ\,\frac{Q^2(S(Q)-1)^2}{S(Q)+b(Q)}\right] \tag{15}$$

where $D_{SE}$ is the Stokes-Einstein (SE) diffusivity in dilute solution. One can define a short time friction constant $\zeta_s = \zeta_0\,[1+\frac{d^3}{36\pi\phi}\int_0^\infty dQ\,\frac{Q^2(S(Q)-1)^2}{S(Q)+b(Q)}]$ where for a colloidal suspension $\zeta_0 = \zeta_{SE}\,g(d)$. In the above equation $\tau_0 \equiv \frac{d^2}{D_0}$, $D_0 = \frac{k_B T}{\zeta_0}$, $Q = kd$, and $b^{-1}(k) = 1 - j_0(k) + 2j_2(k)$ where $j_n(x)$ is the spherical Bessel function of order $n$.

The above NLE based theory only captures the consequences of the local cage on tagged particle hopping. Most recently, the "Elastically Collective NLE" theory (ECNLE) has been developed, widely applied, and quantitatively validated for dense suspensions of hard sphere colloids, cold molecular liquids, and polymer melts [35,36]. It includes a longer range cooperative motion aspect of structural relaxation based on the idea that the fluid surrounding a particle cage must elastically dilate by a small amount (via a spontaneous thermal fluctuation) to accommodate large amplitude hopping. This elastic energy contributes an extra barrier to the activated hopping process given by: $\beta F_{el} = 2\pi K_0 \int_{r_{cage}}^\infty dr\, r^2 \rho g(r) u(r)^2$, where $K_0$ is the harmonic spring constant of the dynamic free energy which sets the energy scale of the elastic barrier, $u(r)$ is the elastic displacement field at a scalar distance $r$ from the cage center $u(r) = \Delta r_{eff}\left(\frac{r_c}{r}\right)^2$, $r > r_c \sim 1.5d$, and the amplitude $\Delta r_{eff} \leq r_L$ the explicit formula for which is given elsewhere [35,36]. Physically, the local and elastic barrier are additive, so the hopping time is modified as a multiplicative factor $e^{\beta F_{el}}$ in the Kramers time as $\beta F_{Total} = \beta F_B + \beta F_{el}$ [35].

The conceptual ideas of ECNLE theory, key length and energy scales, and a representative dynamic free energy are shown in Fig.6 for the Hertzian model. The location of the maximum cage restoring force ($r^*$) obeys $\frac{\partial^2 F_{dyn}(r)}{\partial r^2} = 0$, and the barrier location ($r_B$), jump distance ($\Delta r = r_B - r_{loc}$), and local barrier ($\beta F_B$) are also indicated.

**F. Rheology**

The NLE and ECNLE theories can be extended to treat non-equilibrium materials under deformation. Extensive applications to hard sphere colloids, polymer-colloid depletion systems, polymer glasses, molecular colloids, and nanoparticle gels have been made [12,37-39]. The approach assumes the dominant effect is the direct consequence of applying stress to the material, which leads to an effective force on each particle in a micro-rheological spirit. Technically, a stress ensemble (creep) is adopted to formulate the basic ideas. It is asserted that a macroscopic stress manifests itself locally as a scalar applied force on any tagged particle given by [37]

$$f = ad^2\sigma \tag{16}$$

where $a = \pi/_6\, \phi^{-2/3}$. Stress then modifies the dynamic free energy as [37]

$$\beta F_{dyn}(r,\sigma) = \beta F_{dyn}(r,\sigma = 0) - \beta\ \pi/_6\, \phi^{-\frac{2}{3}} d^2\sigma\, r. \tag{17}$$

External forces are assumed to not modify structural correlations on the *local* length scales dynamically relevant in the theory, nor the short time relaxation process in $\tau_s$. Increasing the applied stress weakens the localizing constraints of the dynamic free energy, and hence reduces the barrier and can mechanically drive a glass-to-liquid transition. At a critical value of stress, called the "absolute yield stress", $\sigma_{y,abs}$, the barrier is completely destroyed, indicating an athermal type of solid-to-liquid transition. With increasing force or stress below its absolute yield value, the localization length grows and the elastic shear modulus decreases continuously. A simple nonlinear elastic mechanical equation-of-state (relevant in practice at times short compared to stress relaxation times) previously adopted implicitly defines strain as [12,37]:

$$\sigma = G'(\sigma)\gamma. \tag{18}$$

This equation can be used to define an "absolute yield strain"

$$\gamma_{y,abs} = \frac{\sigma_{y,abs}}{G'(\sigma_{y,abs})}. \tag{19}$$

Other types of yield strains such as a "dynamic yield strain" can also be defined as the strain at which $G''(\gamma)$ has a maximum within the framework of a one structural relaxation time model which is a function of applied deformation. The nonlinear loss modulus is modeled as [36,38]:

$$G''(\gamma) = G'(\gamma)\frac{(\omega\tau_\alpha(\gamma))^2}{1+(\omega\tau_\alpha(\gamma))^2}. \tag{20}$$

"Mixed" yield strains can also be defined as [12,37]:

$$\gamma_{y,mix} = \frac{\sigma_{y,abs}}{G'(0)}. \tag{21}$$

The stress dependent relaxation time follows from the same Kramers' hopping time expression but where all dynamic free energy quantities are now stress-dependent,

$$\frac{\tau_\alpha(\sigma)}{\tau_s} = 1 + \frac{2\pi}{\sqrt{K_0(\sigma)K_B(\sigma)}} e^{\beta(F_B(\sigma)+F_{el}(\sigma))}. \tag{22}$$

A predictive theory for the full stress-strain response, time-dependent creep, steady shear flow curve, etc. can be constructed [39] but this is beyond the scope of the present work.

## V. Model Calibration, Glassy Shear Modulus, and Collective Structure Predictions

In this section, we employ the microgel model of section IVA to determine the effective volume fraction for our microgel suspensions. We then use this knowledge to perform theoretical calculations of the linear elastic shear modulus and compare to experiment.

### A. Effective Microgel Radius and Volume Fraction in Dense Suspensions

The effective volume fraction ($\phi_{eff} = \frac{\pi}{6}\rho d^3$) depends on concentration via the microgel diameter, $d(c)$. As discussed in section IVA and Figure 5, experiments suggest there are two regimes where the microgel radius first decreases weakly ($R_g \sim c^{-1/6}$) starting at $0.4\, wt\%$ whence $\phi \sim c^{1/2}$ , which then changes beginning at 1.5 $wt\%$ to a stronger shrinkage $R_g \sim c^{-1/3}$ and hence $\phi_{eff} \neq f(c)$. The chosen crossover concentration is motivated by our physical hypothesis that the sharp change of the elastic modulus data in Fig.1B is an indication of a change of the scaling of microgel size with concentration. Figure 7 presents the quantitative model employed for microgel size and effective volume fraction as a function of concentration. The latter ranges from ~0.5 to 0.88. As an independent estimate of the effective volume fraction for our $0.5wt\%$ sample, we have applied our approach to data from literature [40] for a similar microgel system and find it gives $\phi = 0.45 \sim 0.55$ for $c = 0.5wt\%$, consistent with Fig.7.

The one remaining unknown in our model is the dimensionless strength of the Hertzian repulsion, the parameter $E$ in Eq.(1). For simplicity, and to avoid introducing an adjustable function, we assume this is a material constant invariant to concentration. This simplification seems consistent with the very recent simulation study [30] that included the internal polymeric degrees of freedom of a microgel. We can then apply the theory ideas of sections IVA, IVB and IVC to calculate the dynamic elastic shear modulus. We ask the question whether it is possible to theoretically predict the entire set of linear elastic modulus data in both the glassy and soft jamming regimes of Fig.1B based on a single constant value of varying $E$. There is no guarantee the answer is yes.

### B. Linear Elastic Modulus: Theory versus Experiment

The inset of Figure 8 shows model calculations of the dimensionless linear shear modulus, $G'/(k_BT/d^3)$, over a wide range of $E$ values. Recall that the data of Fig.1B in the glassy regime covers almost $\sim 3$ decades of modulus variation. Given the theoretical model calculations and experimental data, this places a constraint on possible values of $E$. Values of $E$ lower than those shown in the inset of Fig.8 cannot possibly account for our observations. Based on these considerations, and visual comparison of the theory and experimental results for the elastic modulus, we choose $E = 30{,}000$ to explore the ability of the theory to account for the entire $G'$ data set. This $E$ value corresponds to a single particle Young's modulus of $Y \approx 1.5\ kPa$ ($\nu = 0.5$), which seems a reasonable value for our lightly self-crosslinked and neutral microgels.

Before quantitatively confronting theory with experiment, we note that the NMCT-based theory of the elastic shear modulus that employs the approximation of Eq.(11) is, of course, not exact. It has been successfully employed to understand how particle and thermodynamic state variables determine dependences and trends of the elastic modulus in diverse colloidal glass and gel forming suspensions [12,37,41] and molecular and polymeric liquids [36,42]. However, concerning the absolute magnitude of the dynamic modulus, multiple previous studies and comparisons with diverse experimental systems (colloids, molecules, polymers) have consistently shown that NMCT quantitatively overestimates particle localization and hence $G'$, which is at least partially likely a consequence of its formulation at the single particle dynamics level [12,36,37,41,42]. Specifically, Eq.(10) has been found to generically overpredict $G'$ by roughly one order of magnitude. Thus, we introduce a numerical 'fudge-factor' to empirically rescale the theoretical result for all microgel concentrations, $G' = 0.1G_{NMCT}$.

To compare theory with experiment, we use the model of Fig.7 for the effective microgel diameter and volume fraction and $\frac{k_B T}{(100nm)^3} = 4.2Pa$ at room $T$. The results are shown in absolute units in the main frame of Fig.8, and the corresponding dimensionless unit comparison in its inset. We first discuss the glassy regime. One sees from the main frame that, rather remarkably and nontrivially, all the experimental data points essentially fall onto the theoretical curve based on using E=30,000. Considering the high uncertainties of the data for the lowest microgel concentration $c = 0.4wt\%$, we have chosen to ignore this data point for the purpose of assessing the quality of the theoretical analysis. The last data point in the glassy regime ($c = 1.5wt\%$) corresponds to $\phi = 0.88$. As discussed in the next section, this is very close to where *structural* "soft jamming" is predicted based on our calculations of the equilibrium structure of the suspension where the volume fraction at which the cage peak of $g(r)$ is a maximum is the metric [43] adopted to operationally define the soft jamming crossover.

The sensitivity of our elastic modulus predictions to the value of $E$ is illustrated in Figure 8. The blue solid curve is for $E = 30{,}000$, while the gray band covers results over the range of $E = 20{,}000\ to\ 40{,}000$. Red and black points show experimental data below and beyond the onset of “soft jamming”. The blue theory curve follows well a power law concentration dependence of $G'(c) \sim c^{5.6}$ in the glassy regime, very similar to experiment. Our calculations agree well with the data in the glassy regime for this relatively narrow range of $E$, but not outside of it.

At concentrations beyond $c = 1.5wt\%$, the effective volume fraction is fixed per the isotropic microgel compression idea discussed in section IVA. Thus, this idea alone, in conjunction with Eq(11), immediately predicts a crossover of $G'$ to a linear growth with concentration since the dynamic shear modulus scales as $G' \sim \frac{k_B T}{d^3} \sim c$ and the ratio $\frac{r_L}{d}$ is a constant if the effective volume fraction is constant. The blue line in Fig.8 beyond the soft jamming onset is the predicted linear $G'(c) \sim c$ dependence, and agrees rather well with the data.

We emphasize that our theoretical analysis in the very high concentration regime is *not* in the spirit of granular jamming and a literal force contact network, nor the idea that the suspension acts as a homogeneous rubber network, scenarios (1) and (2) discussed in section IVA. Effectively we retain a discrete particle picture with stresses of interparticle Brownian origin due to caging. The "soft jamming crossover" in Fig.1B is thus interpreted as a consequence of the particle size decreasing as the 1/3 root of concentration, which implies a constant effective volume fraction, but a shear stress scale of $kT/R^3$ that grows linearly with concentration.

### C. Predicted Intermolecular and Collective Equilibrium Structure

Given the apparent success of our single microgel model for predicting the dynamic shear modulus of our system, we now use it to explore its consequences for measurable aspects of equilibrium structure. Figure 9 shows predictions for the real and Fourier space pair structure using the "best fit" value of $E = 30{,}000$ over a wide range of volume fractions. Figure 10 quantifies various metrics of the structural correlations in wave-vector and real space. Figures 9 and 10 show that as the effective volume fraction grows, the "contact value" (local maximum) of $g(r)$ (crucial for transmitting repulsive forces between microgels) first grows but then goes through a maximum at a volume fraction of $\approx 0.85$ and decreases beyond that; there is also a splitting of the second peak. This behavior was previously found theoretically [12], and in the simulations and experiments of Liu, Yodh and coworkers [43]. The maximum of the contact value was taken to be an empirical measure of the "soft jamming crossover" by the latter workers. The emergence of a split second peak occurs at essentially the same value of volume fraction $\phi_J \approx 0.85$ as where the first peak is a maximum, which is far beyond $\phi_{rcp} = 0.64$ of jammed hard sphere suspensions. On the other hand, $S(k)$ shows a monotonic growth of cage coherence defined as the amplitude of the first peak of the static structure factor, $S(k^*)$, with increasing volume fraction.

The inset of Figure 10 presents calculations of the zero wave-vector value of $S(k)$, $S_0 = \rho k_B T \kappa_T$ , which is a dimensionless measure of the osmotic compressibility of the suspension. It decreases strongly and monotonically with increasing volume fraction. Integration over concentration of the inverse of this quantity provides the osmotic pressure [44]. In principle the results of Figures 9 and 10 can be tested via new experiments on our microgel samples such as confocal imaging, scattering, and thermodynamic measurements. We now use the obtained structural knowledge to make further dynamical and rheological predictions in the next section.

**VI. Dynamics and Rheology Predictions and Comparison to Experiment**

` To convert our dimensionless theoretical time scales into absolute time scales relevant to our system, we estimate the *short* relaxation time of Eq(15) and find $\tau_s \geq 200\ s$ since the peak value of $g(r)$ obeys $g(d) \geq 4$, and the factor in square brackets in Eq(14) is ~100 at the high effective volume fractions of interest. This estimate also employed the experimental particle radius, the SE diffusivity $D_{SE} = \frac{k_B T}{6\pi\eta R}$ , a water viscosity of $10^{-3} N.s/m^2$, and $T = 10°\text{C}$ . We note $\tau = \frac{d^2}{D_{SE}} = 0.4s$ for a $d = 550$nm diameter particle.

### A. Quiescent Relaxation

To test if our theoretical approach is consistent with the nearly flat frequency dependence of the shear modulus observed experimentally (Fig.1A), we consider a simple Maxwell model defined as

$$G'(\omega) = G' \frac{(\omega\tau_\alpha)^2}{1+(\omega\tau_\alpha)^2} \tag{23}$$

where $G'$ is given by Eq(11). A flat frequency dependence requires $\omega\tau_\alpha \geq 1$. In the experiments the lowest frequency probed is ~$10^{-2}\ rad.s^{-1}$ . Using this and our calculation of the short time scale $\tau_s \geq 200\ s$, we find $\omega\tau_s \geq 2$. Indeed, the actual structural relaxation time, estimated here as the Kramers time, is much larger than $\tau_s$. Since we interpret in a Maxwell model spirit the structural and longest stress relaxation times to be essentially the same to leading order, the inequality $\omega\tau_\alpha >> 1$ applies and thus the dynamic theory is consistent with the observation of no terminal flow on the experimental time scale under quiescent conditions.

As discussed in section IVE, the dynamic free energy has several key length scales per Fig.6. Figure 11 shows examples using $E = 30{,}000$. All length scales are 1-2 decades smaller than the particle size. The transient localization length ($r_{loc}$) and location of maximum force ($r^*$) monotonically decrease (initially strongly) with volume fraction, and then tend to saturate as the soft jamming point is approached. The jump distance grows monotonically. Our predictions of localization length can potentially be tested using confocal microscopy or simulations.

Calculations of the local cage, collective elastic, and total barriers discussed in section IVE are shown in Fig.12a. All grow monotonically and strongly with volume fraction over the range shown. The collective elastic barrier increases more strongly with concentration, as also true for hard spheres and other glass forming liquids [35,42]. The elastic and local barriers cross at a much higher volume fraction than for hard spheres, and the crossing point decreases as $E$ grows (not shown).

### B. Nonlinear Response

With increasing deformation or stress, both dynamical barriers decrease and the structural relaxation time strongly decreases. Figure 12b shows this is an extremely dramatic effect for five different concentrations below the soft jamming threshold. The last point in each plot corresponds to when the localized form of the dynamic free energy is first destroyed (and hence the total barrier vanishes), which signals the absolute yield stress.

Figure 12b can also be used to operationally define a dynamic yield stress in the spirit of a mechanically-driven glass to liquid transition. Typically, the kinetic criterion used is set by the maximum experimental observation time. For example, the dynamic yield stress could correspond to the stress value when $\tau_\alpha = 10^x$ s where $x \sim 2 - 4$. But here we choose to do a simpler analysis by defining [12,37] a dynamic yield stress as $\sigma_{y,dyn} = \gamma_{y,dyn} \times G'\ (\gamma_{y,dyn})$ in analogy with Eq.(18), where $\gamma_{y,dyn}$ is the dynamic yield strain defined at the maximum of the strain dependent loss modulus, $G''(\gamma)$ of Eq(20). Another way of defining yield strain is where the strain dependent storage and loss moduli cross, $G''(\gamma) = G'(\gamma)$. Within the simple nonlinear Maxwell model framework of Eq. (20), these two definitions are the same. Experimentally, these two criteria may be different (Figure S5). We take the peak in $G''$ as the dynamic yield strain and the crossover as the absolute yield strain for comparison to theory. Figure 13 presents our theoretical results for the dynamic and absolute yield stresses and strains, and compares them in a no adjustable parameter manner to experiment.

Figure 13 shows rather good agreement between different theoretical measures of the yield stress (smooth curves) and experimental data analyzed in 3 different ways (data points) in both the glassy and soft jamming regimes (except for the lowest concentration sample for which the data is most uncertain). The inset compares yield strains from theory and experiment. Overall, the agreement is good in the glassy regime where the system has yield strains of modest magnitude, ~ $10 - 20$ %. Agreement between theory and experiment is not very good beyond the putative "soft jamming" crossover. While theory predicts $\gamma_y \sim c^0$, experiment suggests a strong yield strain dependence on concentration at very high concentrations, leading to a large yield strain value of $\sim 72\%$ for the 9 $wt\%$ sample. This is much larger than the theoretical predictions and may reflect the arbitrariness of defining yield strains from real experimental data. Using a different definition, the mixed yield strain (defined in Eq. (21)) evaluated using our experimental data as the ratio of the HB yield stress to plateau modulus ($\gamma_y^{mix} = \sigma_y^{HB}/G_0$), results in a nearly constant yield strain $\gamma_y^{mix} \sim c^0$ in the highest concentration regime.

## VII. Summary and Conclusions

We have presented an integrated experimental and quantitative theoretical study of the linear and nonlinear rheology of self-crosslinked, neutral pNIPAM microgel suspensions at low temperatures where they repel. An exceptionally wide range of concentrations were studied that span the fluid, glassy and so-called "soft jammed" regimes. In the intermediate glassy regime, we measured over 3 orders of magnitude an apparent power law dependence of the elastic shear modulus on concentration, $G' \sim c^{5.64}$. This variation appears distinct compared to prior studies of crosslinked ionic microgel suspensions. At high enough concentrations, there is a rather sharp crossover to a nearly linear growth of the dynamic shear modulus. To theoretically understand these quiescent observations within a single framework we constructed a minimalist model of single microgel size as a function of concentration that includes steric de-swelling effects which differ in the so-called glassy and highest concentration or soft jammed regimes. Using a Hertzian repulsion interparticle potential and a suite of statistical mechanical theories, we made quantitative predictions for the microgel collective structure, dynamic localization length, and elastic shear modulus. Based on a constant Hertz repulsion strength parameter ($E$), determined by requiring the theory to reproduce the measured elastic modulus over the entire concentration regime studied, we demonstrated good agreement between theory and experiment for $E \approx 30{,}000$. Experimentally testable predictions were made for the structure of the suspensions.

We also measured several nonlinear rheological properties with a focus on the yield stress and strain. Again significant differences of our data compared to published studies using crosslinked ionic microgels were found [5,8-11]. A theoretical analysis was also performed (now with no adjustable parameters) to predict the structural relaxation time in equilibrium, how it changes under deformation, and the yield stress and strain as a function of microgel concentration. Reasonable agreement with our observations was obtained. To the best of our knowledge, this is the first theoretical attempt to quantitatively understand structure, quiescent relaxation and shear elasticity, and yielding of dense microgel suspensions using microscopic force based methods that include activated hopping processes.

We expect the ideas and approach presented here will be useful for other realizations of microgel suspensions based on different chemistries and also other types of soft polymeric particles in the core-shell family. A key input to the modeling is knowledge of the interparticle pair potential and the microgel size and effective volume fraction as a function of concentration. Given these, the statistical mechanical theories discussed in this article can be employed to predict packing structure in real and Fourier space, the shear elastic modulus, structural relaxation time, and nonlinear rheological properties. Our integrated experimental-theoretical approach will be applied in a future article to study how heating induced changes of microgel size and stickiness impact linear and nonlinear viscoelasticity.

**Acknowledgement.** This work was performed at the University of Illinois and supported by DOE-BES under Grant No. DE-FG02-07ER46471 administered through the Frederick Seitz Materials Research Laboratory.

## Figures

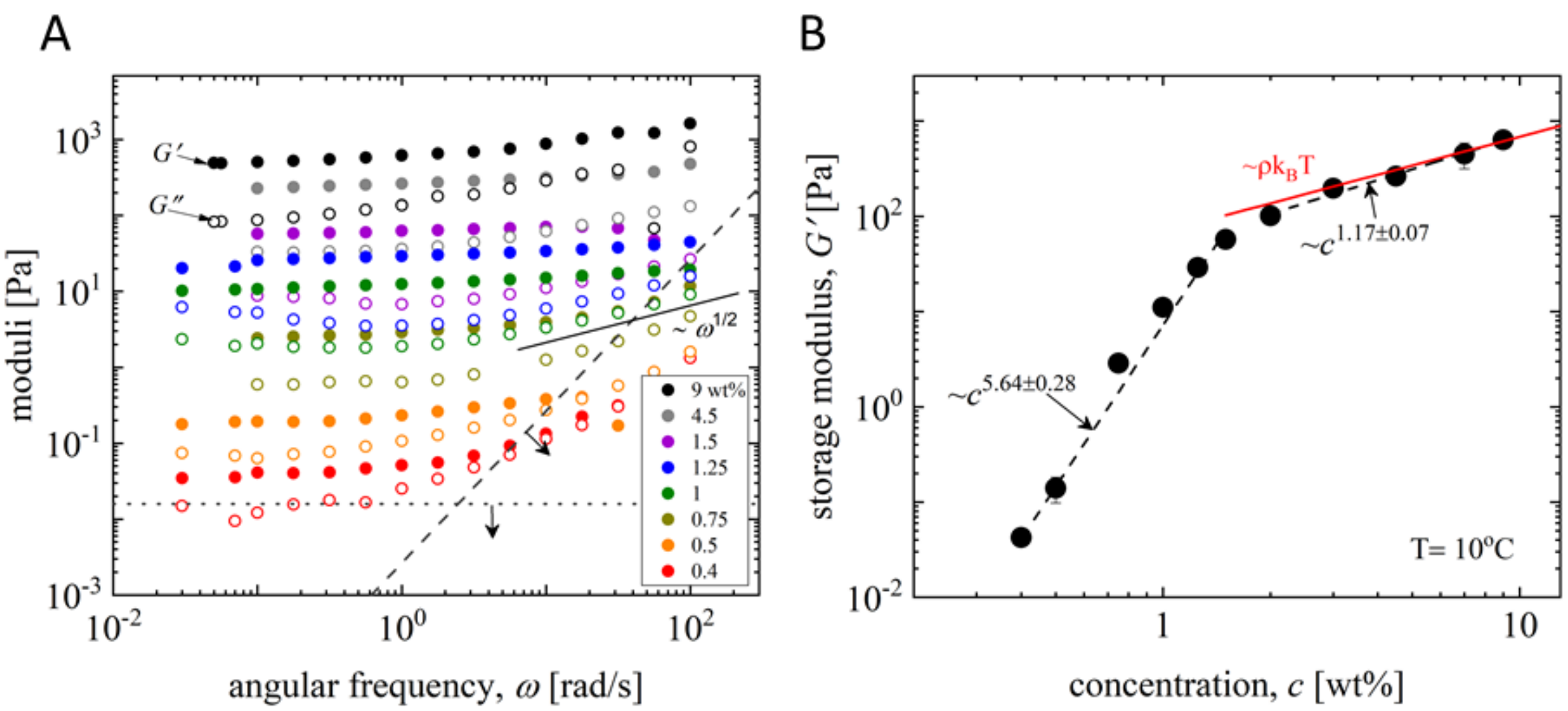

**Figure 1** - Linear rheological response ($G'$ closed symbols, $G''$ open symbols) of the neutral, self-crosslinked microgel suspensions. (A) frequency dependence at $\gamma_0 = 1\%$. Suspensions at $c > 0.4\ wt\%$ do not flow on the longest probed time scales ($\sim 100\ s$). Experimental limits shown by the dotted horizontal line (minimum torque limit) and the dashed line (instrument inertia limit) following [17]. (B) Concentration dependence of linear storage modulus, $G'$. For low concentrations ($c < 1.5\ wt\%$), $G'$ varies over 3 orders of magnitude and roughly follows a power law concentration dependence, $G' \sim c^{5.68 \pm 0.28}$. Above $c = 1.5\ wt\%$, the concentration dependence changes to a roughly linear relation, $G' \sim c$. The red line shows a fit using the classic rubber elasticity model discussed in the text.

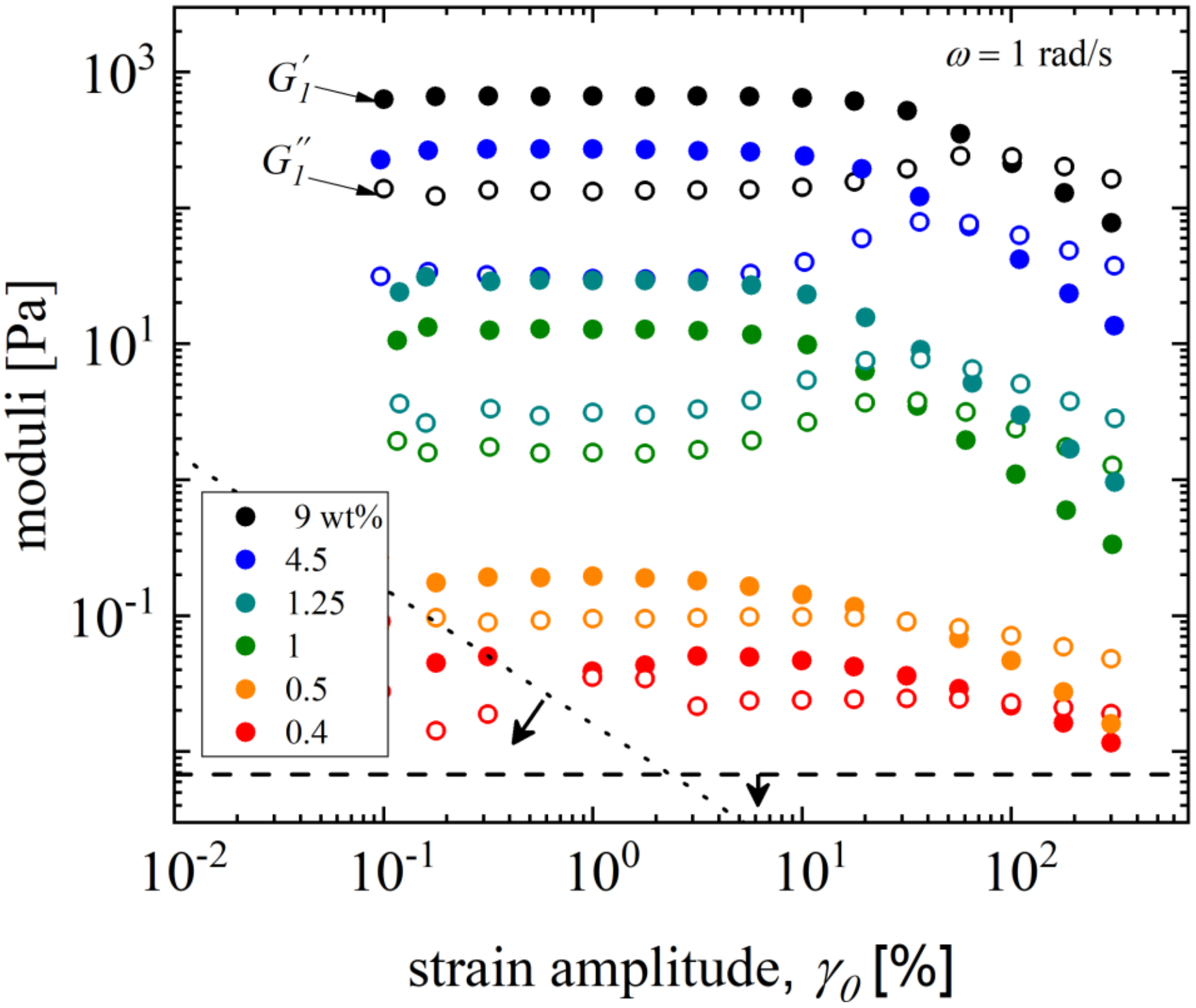

**Figure 2** – Nonlinear viscoelastic moduli (first harmonic $G_1'$ closed symbols, $G_1''$ open symbols) measured at varying strain amplitudes at a fixed frequency $\omega$ = 1 rad/s. At low strains, the response is predominantly elastic, $G_1' > G_1''$ and $G' \sim$ constant. Beyond the linear regime, $G_1'$ monotonically decreases, while $G_1''$ achieves a maximum value as the material undergoes yielding. With further increase in strain, suspensions at all concentrations have a dominant liquid-like response, with both $G_1'$ and $G_1''$ showing a monotonic decrease and $G_1' < G_1''$. The dotted line shows the minimum torque limit of the instrument and the dashed line shows the instrument inertia limit [17].

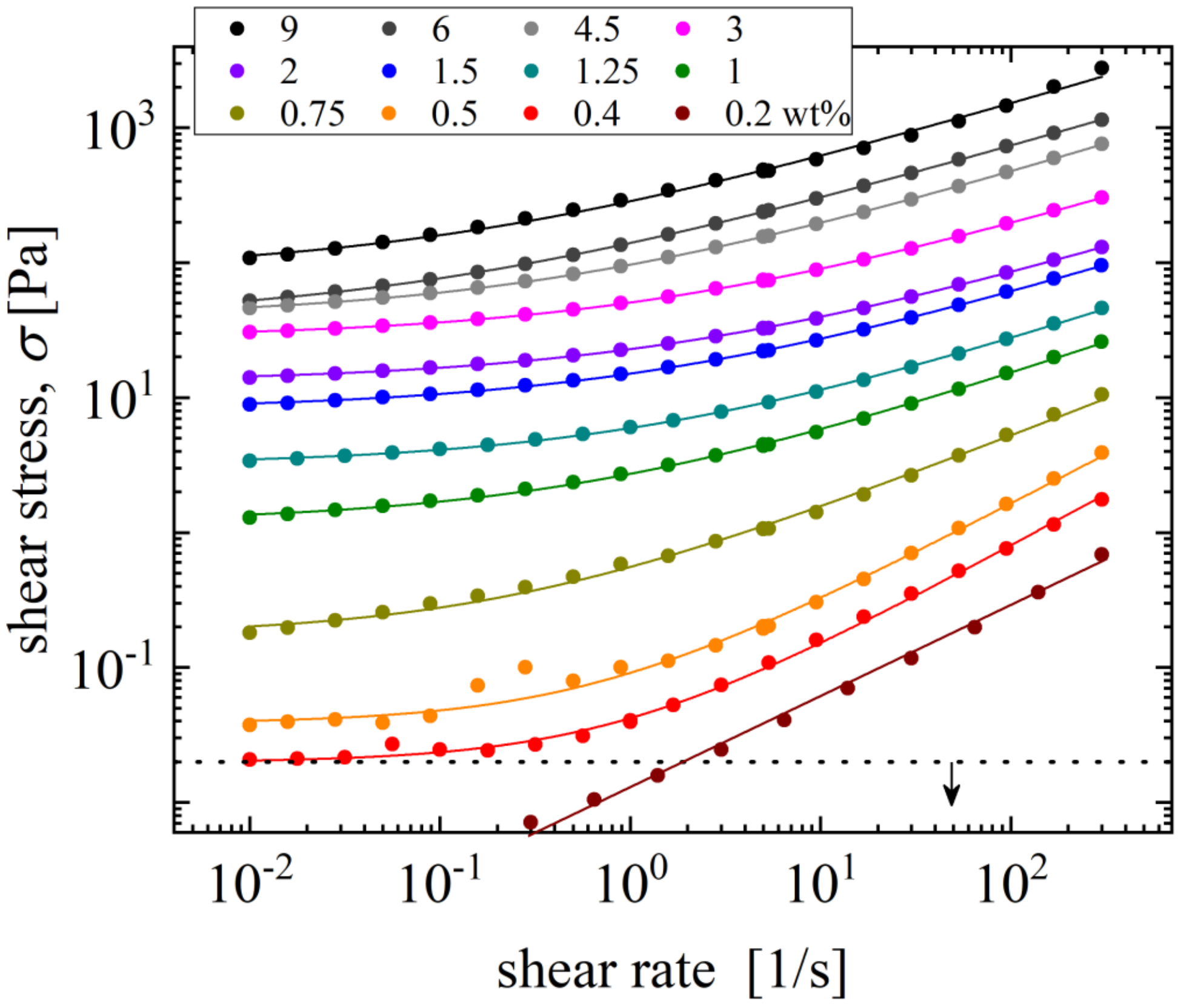

**Figure 3** - Steady state shear flow curves for various suspension concentrations. For $c \geq 0.4\ wt\%$, all suspensions show an apparent yield stress response, achieving a near plateau at low shear rates. For $c < 0.4\ wt\%$, the response closely resembles a shear thinning fluid (power law stress-rate scaling with an apparent exponent smaller than 1) in the range of shear rates probed. The solid curves are the Herschel-Bulkley model fits, $\sigma(\dot{\gamma}) = \sigma_y^{HB}\left(1 + \left(\frac{\dot{\gamma}}{\dot{\gamma}_c}\right)^n\right)$ (Eq.(1)). The dotted horizontal line shows the minimum torque limit of the instrument [17].

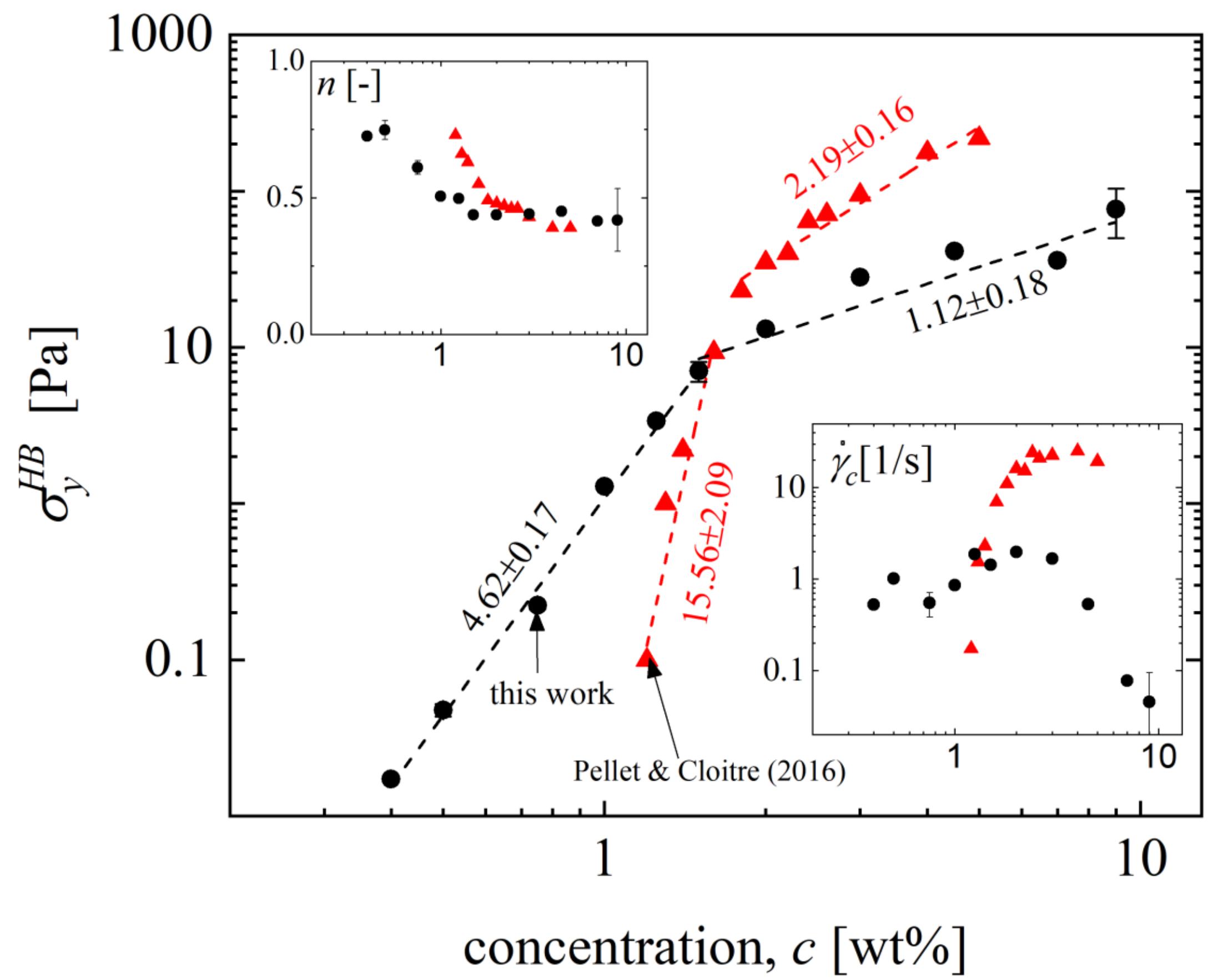

**Figure 4** – Concentration dependence of the Herschel-Bulkley model fit parameters, Eq.(1), for our neutral microgel suspensions (black circles, from data in Fig.3). Data for the ionic microgel suspensions of ref. [11] are shown as red triangles. Power-law scaling exponents are indicated for each fit line. (Inset) Corresponding characteristic shear rate data determined as defined below Eq(11).

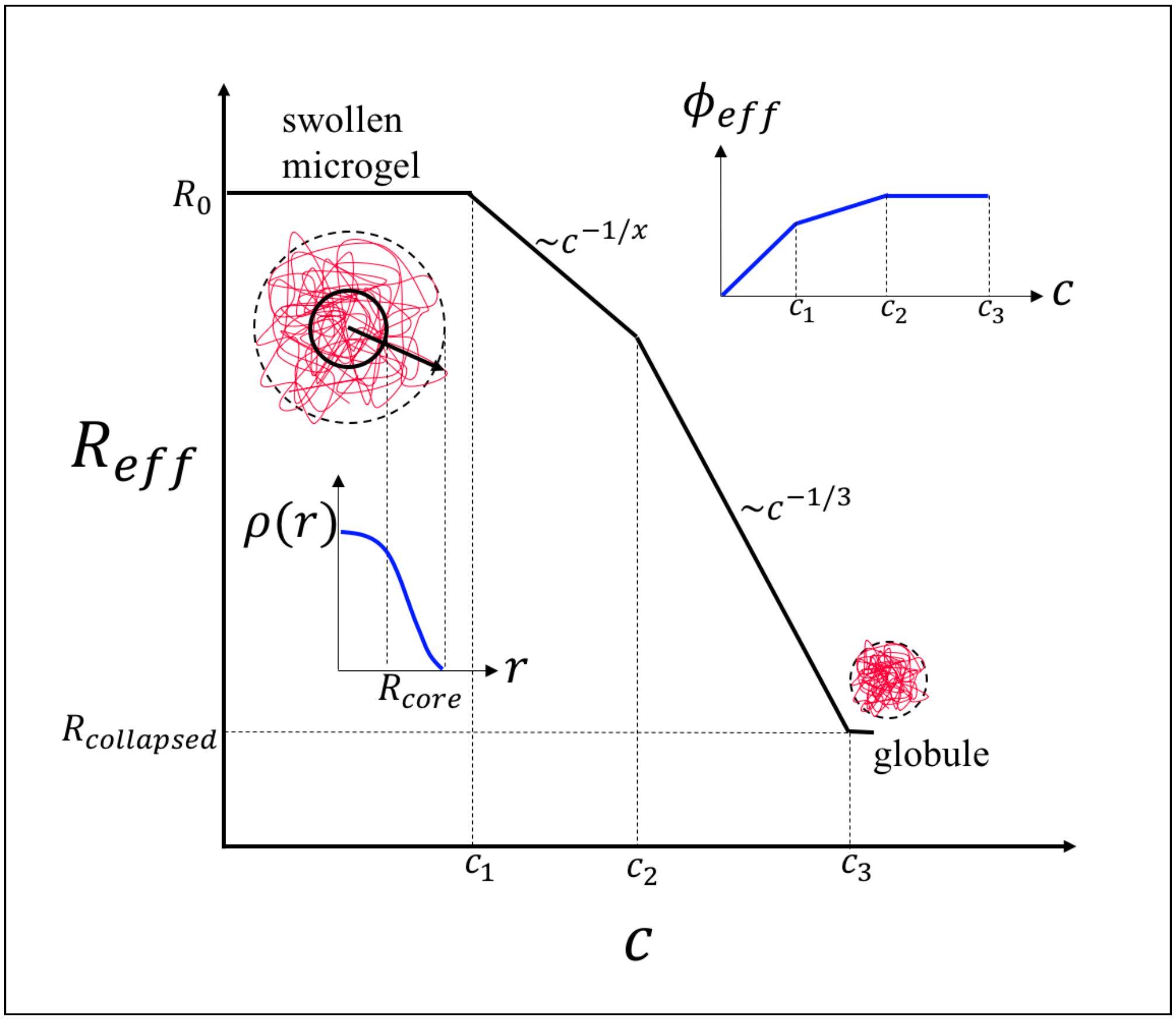

**Figure 5 -** Schematic of our model for microgel radius as a function of concentration. In principle, there can be four regimes. At low concentration, the size is fixed at its $c \to 0$ dilute limit value as measured by DLS. Two intermediate regimes have different concentration dependences in the glassy and "soft jammed" regimes which we envision as physically indicating first compression of the corona and then stronger shrinkage of the core due to interparticle steric repulsions. The final, perhaps not observable, regime is when the core is maximally compressed and microgel size saturates.

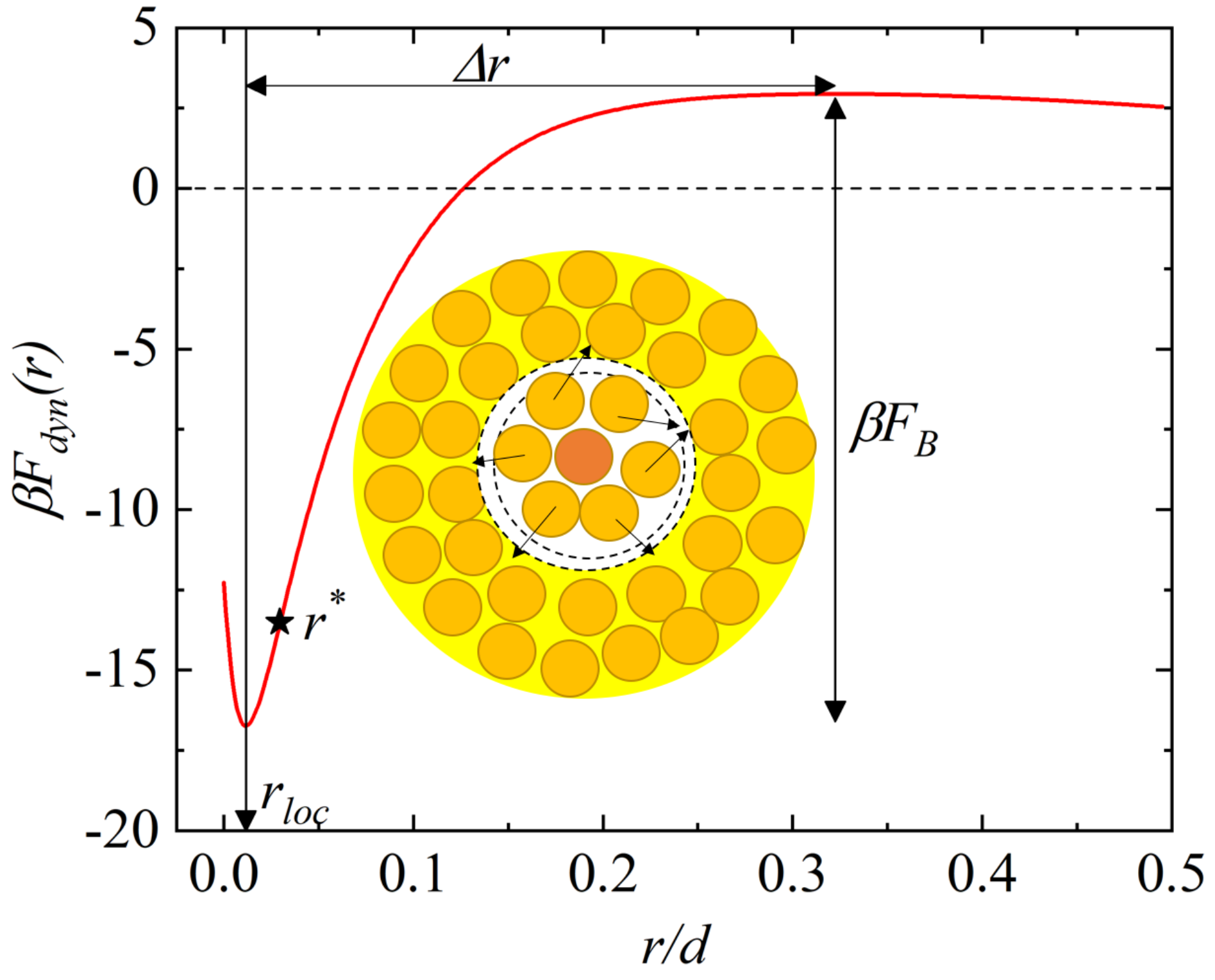

**Figure 6** - A representative plot of the dynamic free energy in thermal energy units as a function of dimensionless single particle displacement from its initial position for a dense suspension. Here $\phi = 0.70$ and $E = 30{,}000$, with all important length scales and the cage local barrier height indicated. The local minimum of the dynamic free energy, $r_{loc}$ , defines the transient localization length, $r = r^*$ is the particle displacement where the cage restoring force is a maximum, and the particle hop or jump distance is $\Delta r$. The schematic indicates a tagged particle at the center of a cage composed of its nearest neighbors, all of which undergo large amplitude hops. To allow the latter, particles outside the cage region undergo a long-range collective elastic radial dilational displacement of small amplitude which results in an elastic contribution to the total dynamic activation barrier.

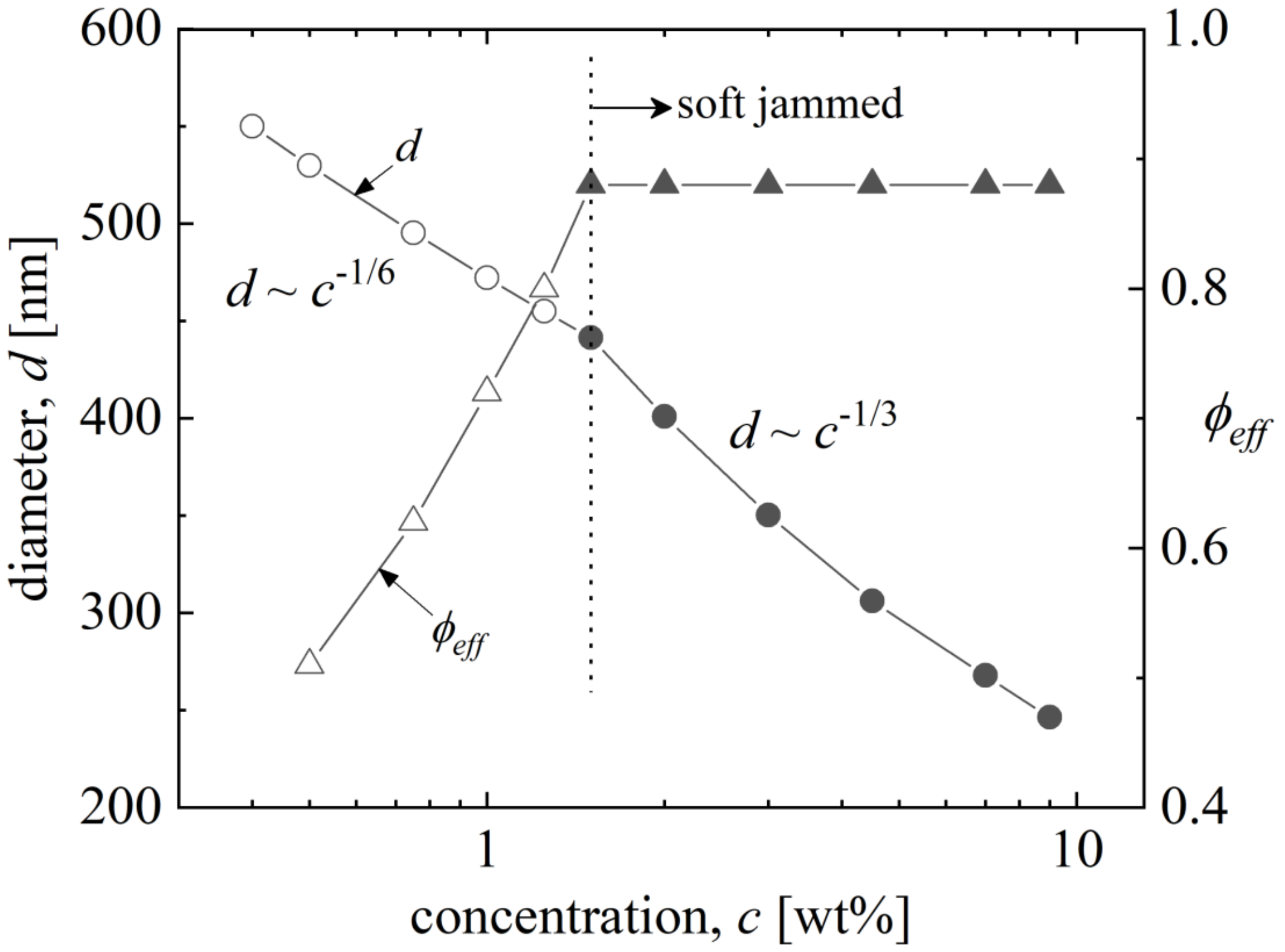

**Figure 7** – Quantitative model employed for the microgel diameter (circles) and effective volume fraction (triangles) as a function of concentration (i.e., quantitative realization of the schematic of Fig. 5). Open symbols indicate the glassy regime while solid symbols indicate the "soft jamming" regime. Here $d = 550nm$ in dilute solution and we assume microgel compression starts at 0.4 $wt\%$.

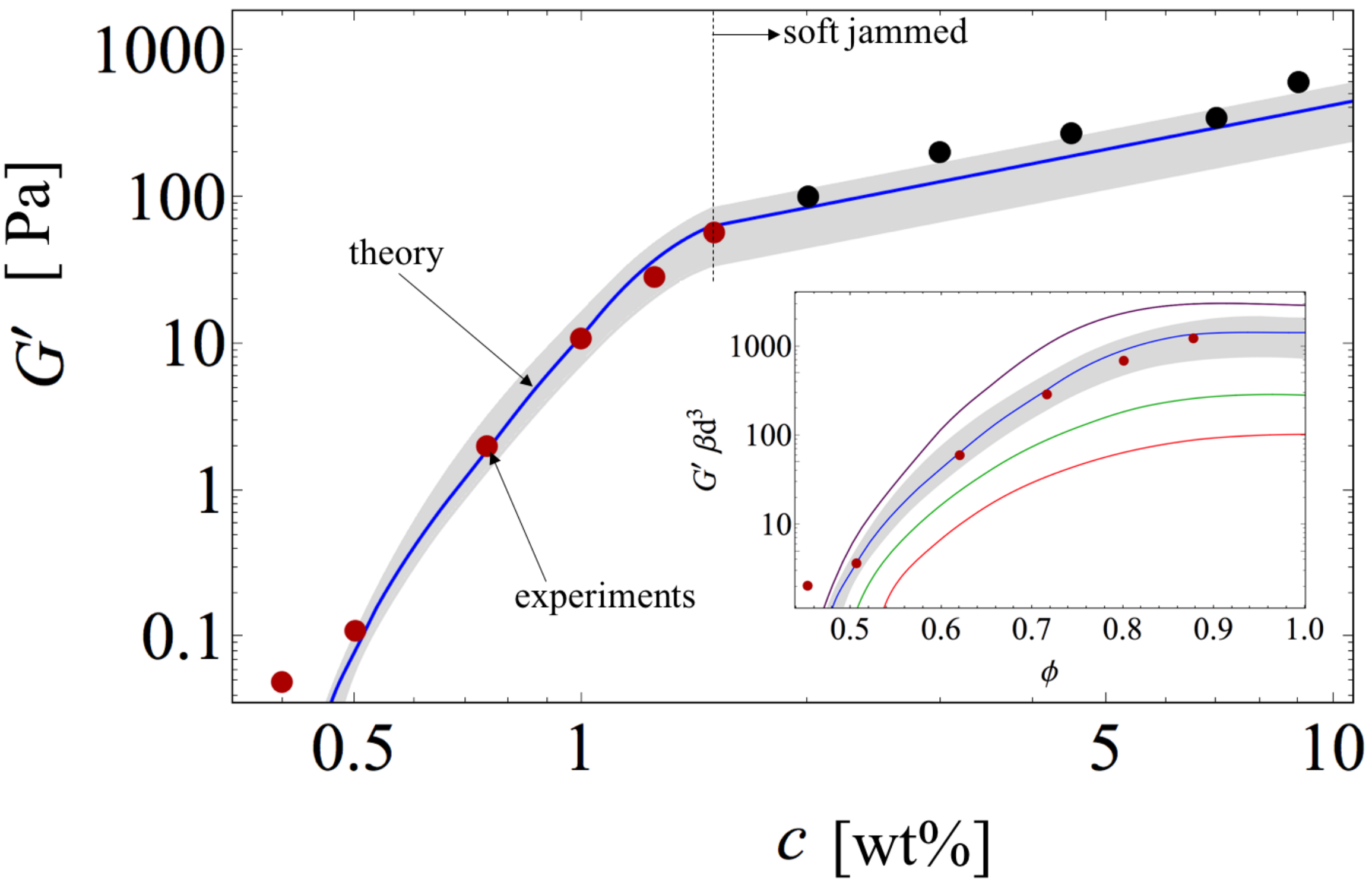

**Figure 8 -** Linear elastic shear modulus in Pascals as a function of concentration. Points indicate experimental data and curves are theoretical calculations using $E = 30{,}000$. Beyond $c = 1.5\ wt\%$, volume fraction is constant and $G' \sim c$ , which agrees well with the experimental results. (Inset) Dimensionless modulus versus volume fraction $\phi$ for $E =$ 5000, 10,000, $30{,}000\ and\ 10^5\ (bottom\ to\ top)$. At high $\phi$ beyond soft jamming, the theoretical $G'$ results tend to saturate or very weakly decrease, trends that are consistent with previous findings for soft microgel potentials [12]. After the last experimental data point in inset, the volume fraction of the system is essentially constant as described in Figure 7. The gray bands in the main frame and inset indicate the range of variation of the predicted elastic modulus as the repulsion strength in the Hertzian potential varies over the range of $E = 20{,}000\ to\ 40{,}000$.

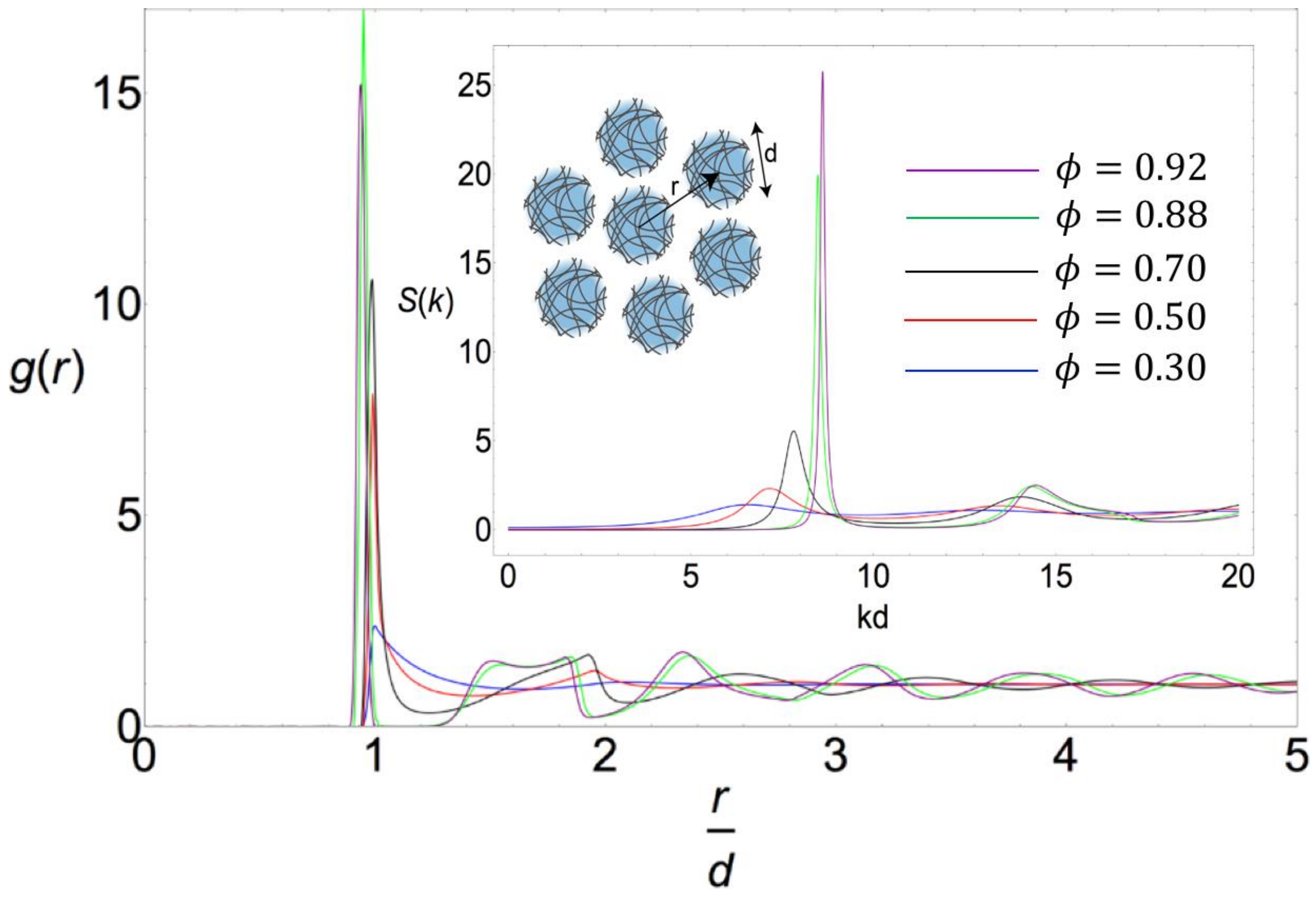

**Figure 9** - Equilibrium pair correlation function as a function of reduced interparticle separation for a fixed repulsion strength of $E = 30{,}000$ over a wide range of indicated volume fractions. (Inset) Static collective structure factor, $S(k)$, for the same value of $E$ and volume fractions. The cartoon shows soft microgels in a transiently kinetically arrested state which are modeled here as Hertzian elastic spheres.

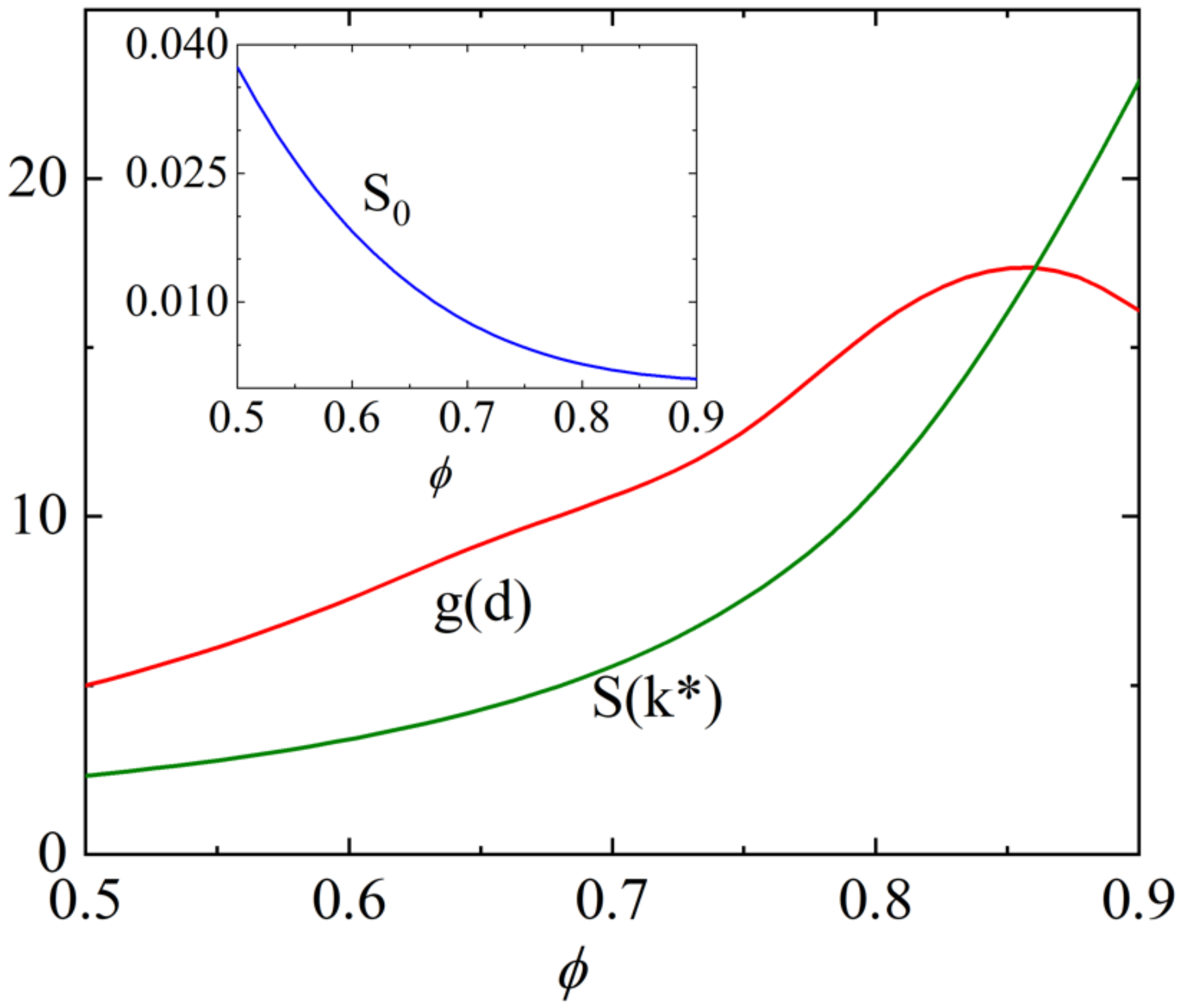

**Figure 10.** Characteristic structural features as a function of volume fraction $\phi$ for Hertzian spheres at a fixed repulsion strength of $E = 30{,}000$. Amplitude of the first peak of $g(r)$, denoted as $g(d)$, is a measure of the degree of real space short range order between nearest neighbors in the liquid. Amplitude of the first peak of the collective static structure factor as defined in section VC, $S(k*)$, which quantifies the collective coherence of cage packing associated with the nearest neighbors. (Inset) Zero wave-vector value of the collective static structure factor, $S_0 \equiv S(k = 0) = \rho k_B T \kappa_T$ , which is a dimensionless osmotic compressibility.

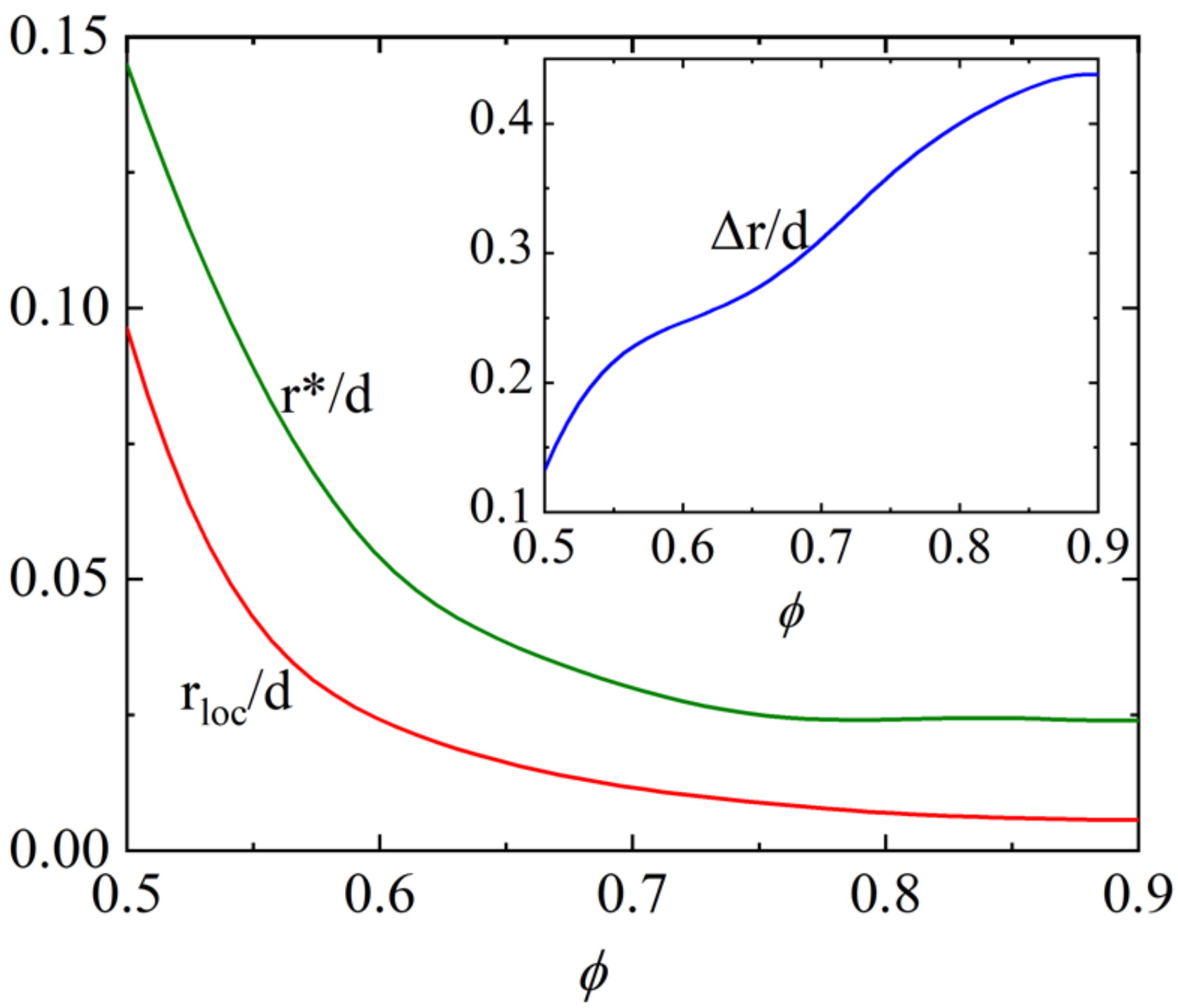

**Figure 11.** Characteristic length scales of the dynamic free energy (c.f. Fig. 6) as a function of volume fraction for fixed $E = 30{,}000$. Dimensionless dynamic localization length, $r_{loc}/d$ (red), and location of maximum cage restoring force, $r*/d$ (green). (Inset) Particle jump distance, $\Delta r = r_B - r_{loc}$ .

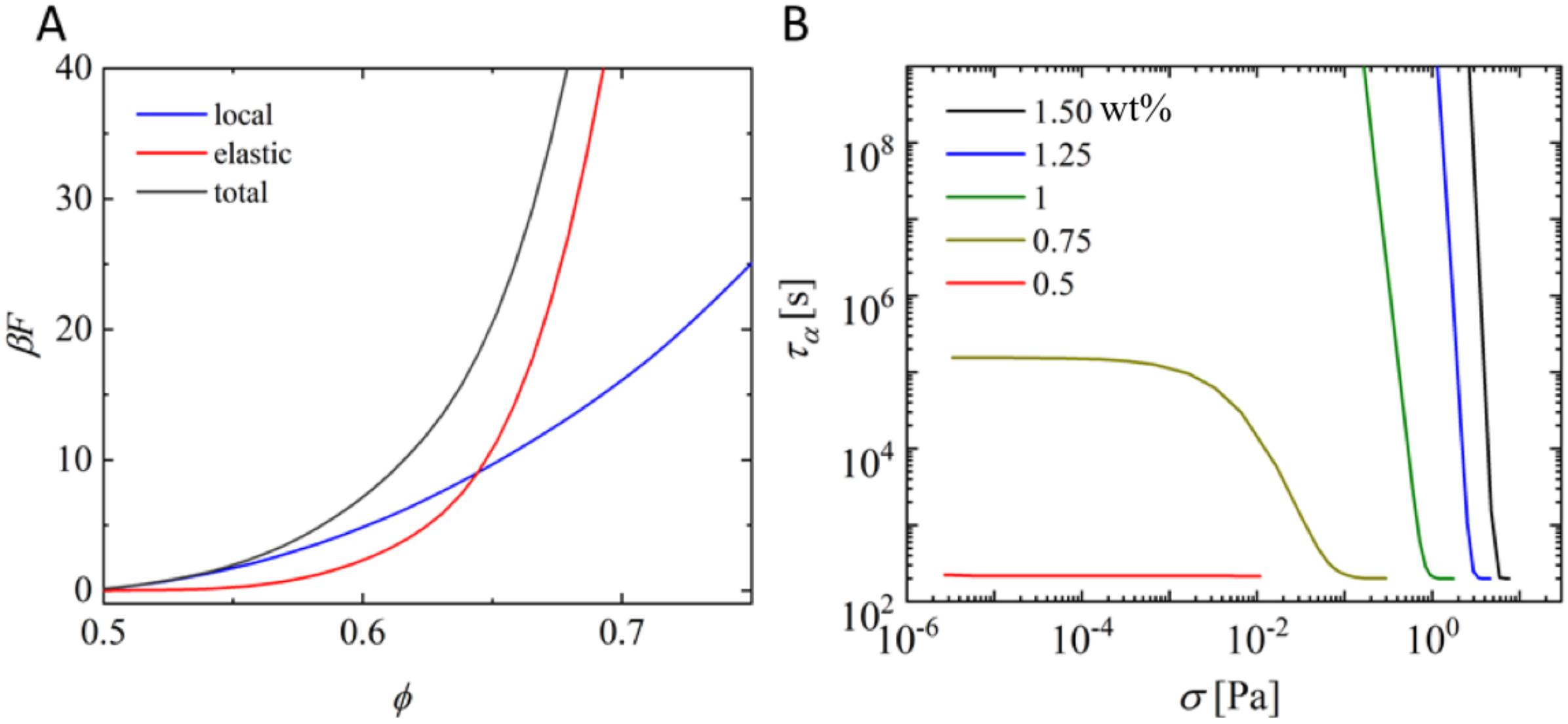

**Figure 12.** . (A) Dimensionless dynamic free energy barriers (c.f. Fig.6) for $E = 30{,}000$. The local, elastic, and total dynamic barriers discussed and defined in section IVE are shown as a function of volume fraction. (B) Alpha relaxation time (in seconds) for five microgel concentrations in $wt\%$ as a function of stress in Pascals.

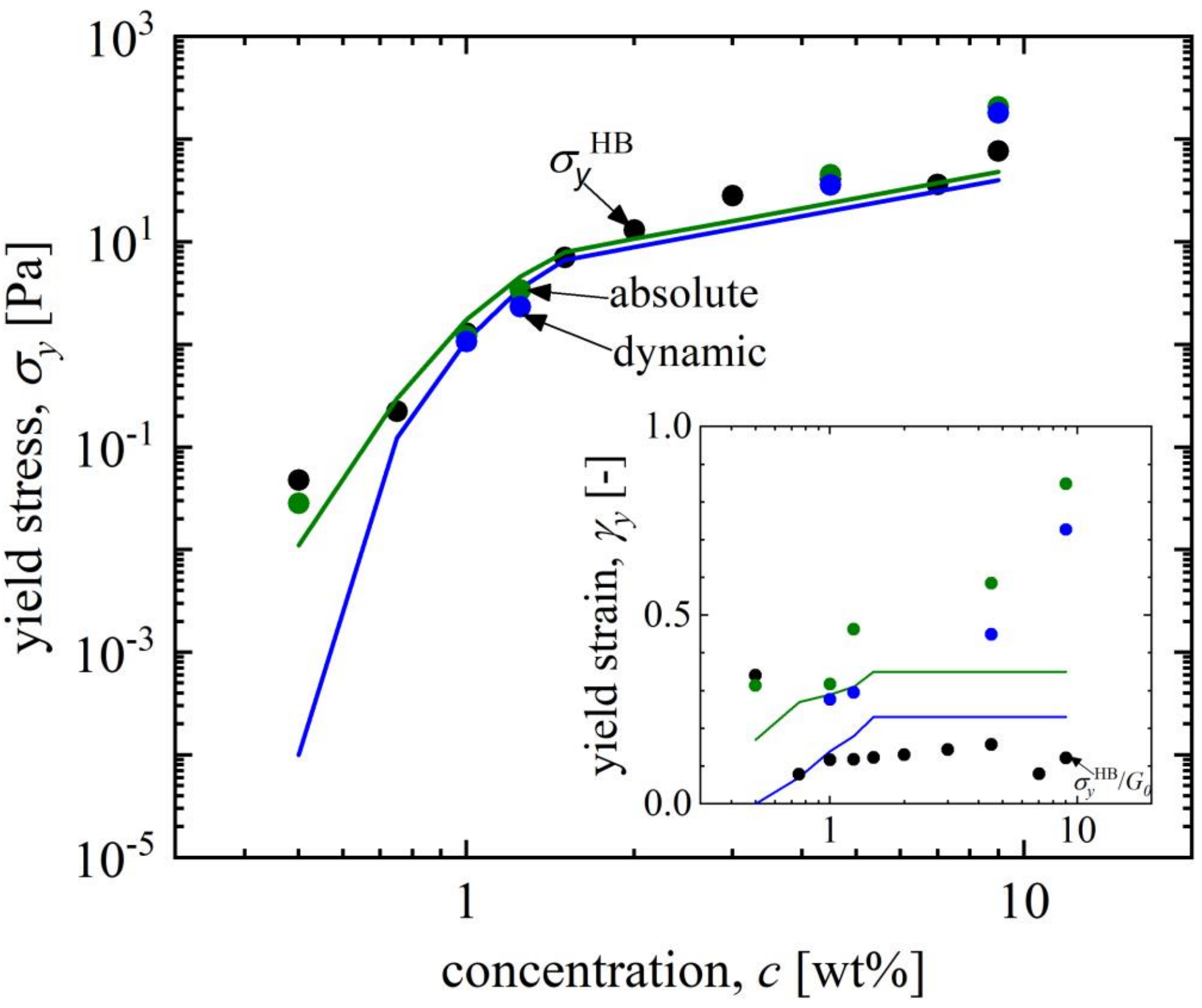

**Figure 13.** Comparison of the yield stress and yield strain from experiment (symbols) and theory with no additional fit parameters (solid curves). Experimental Hershel-Buckley (black), dynamic (blue), and absolute (green) yield stresses as defined in Sec.VI B (from data in Fig. 2, Fig. 3 and Fig. S5). (Inset) Experimental yield strain values (points) and the predicted theoretical dynamic and absolute yield strains as defined in Sec.VI B. These theoretical results are based on the parameters deduced by aligning theory and experiment for the linear shear modulus and involve no horizontal or vertical shifts.

## Supplementary Material

The lightly cross-linked monodisperse PNIPAM microgels were prepared by the surfactant-free emulsion polymerization (SFEP) method [14]. $100\ ml$ of Type I water ($18.2\ M\Omega\ cm$) was filtered through a $0.2\ \mu m$ Acrodisc syringe filter. Then, $146\ mM$ ($1.65\ g$) of N-isopropylacrylamide (NIPAM, 99 %, Acros) monomer was dissolved in filtered water. The monomer solution was again filtered through a $0.2\ \mu m$ Acrodisc syringe filter into a 3-neck round bottom flask. The solution was stirred at $500\ rpm$, purged with nitrogen, and heated to 68$^{o}$C in a temperature-controlled oil bath until the temperature of the solution became stable (1 hour typically). We then injected a solution of $2.8\ mM$ ($80\ mg$) potassium peroxodisulfate (KPS, 99 %+, Sigma-Aldrich) dissolved in 1 ml of the pre-filtered Type 1 water through a $0.2\ \mu m$ Acrodisc syringe filter to initiate the polymerization. The mixture was left to react under continuous stirring at 500 rpm in nitrogen atmosphere overnight. After the polymerization, the solution was cooled down to the room temperature and filtered with a glass wool five times to remove large particulates. The microgel particles were then thoroughly purified via five cycles of a centrifuge/dispersion process. The centrifugation was done at 15000 xg of relative centrifugal force (RCF), and the dispersion was enabled by a mixed process of the ultrasonication followed by the magnetic stirring. The cleaned particles were then lyophilized for further characterization.

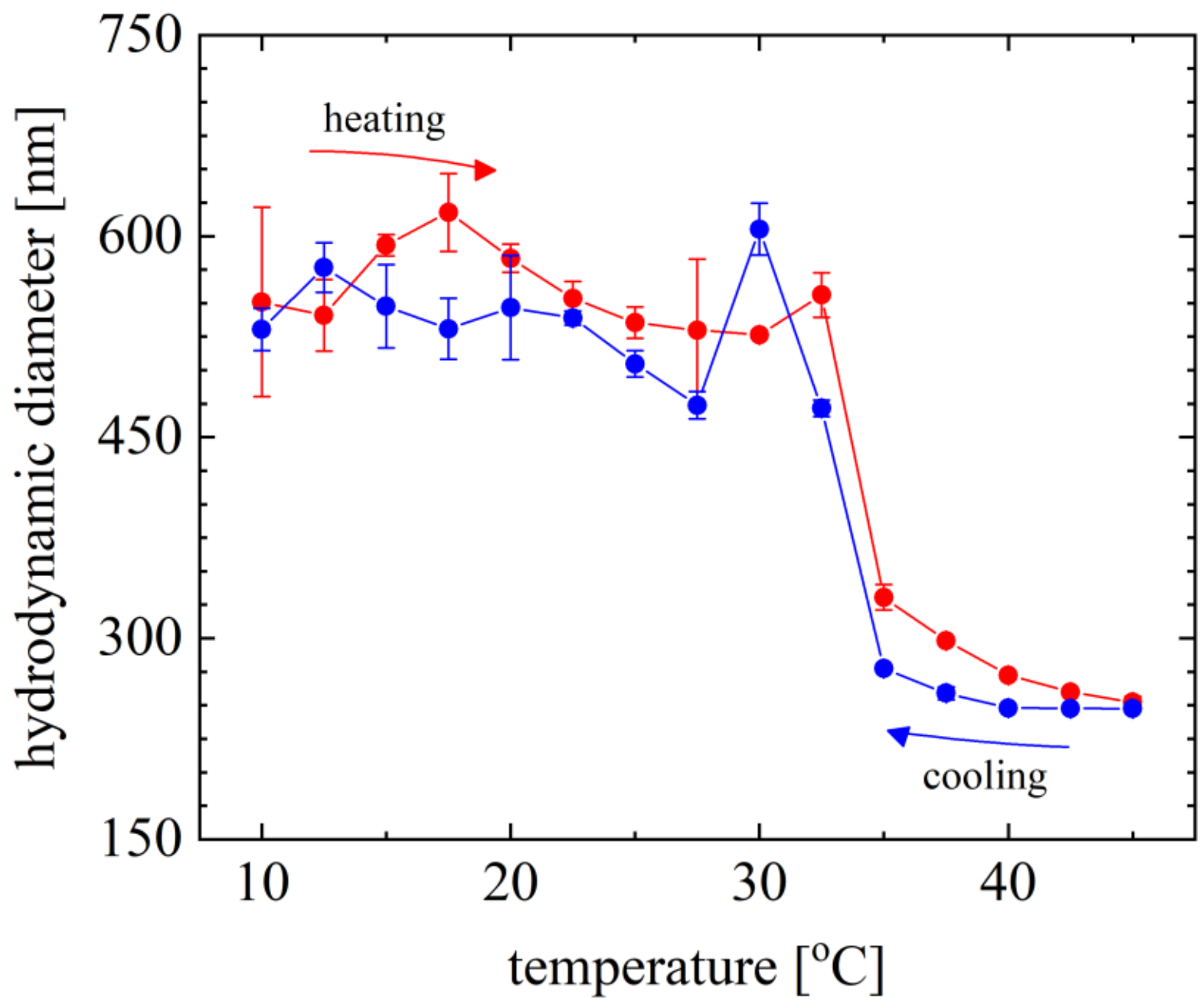

**Figure S1** -Temperature dependence of the hydrodynamic diameter in the low concentration limit ($0.04\ wt\%$) of neutral microgels measured via DLS. As temperature increases in the region $T = 10 - 32$℃, there is a weak roughly linear decrease of the average hydrodynamic diameter. As the lower critical solution temperature (LCST) of pNIPAM microgels is crossed, microgels become hydrophobic and undergo massive deswelling. We observe a hysteresis in the diameter versus temperature plot, as the system is heated and cooled, presumably due to lower water retention of individual microgel particles as they are cooled.

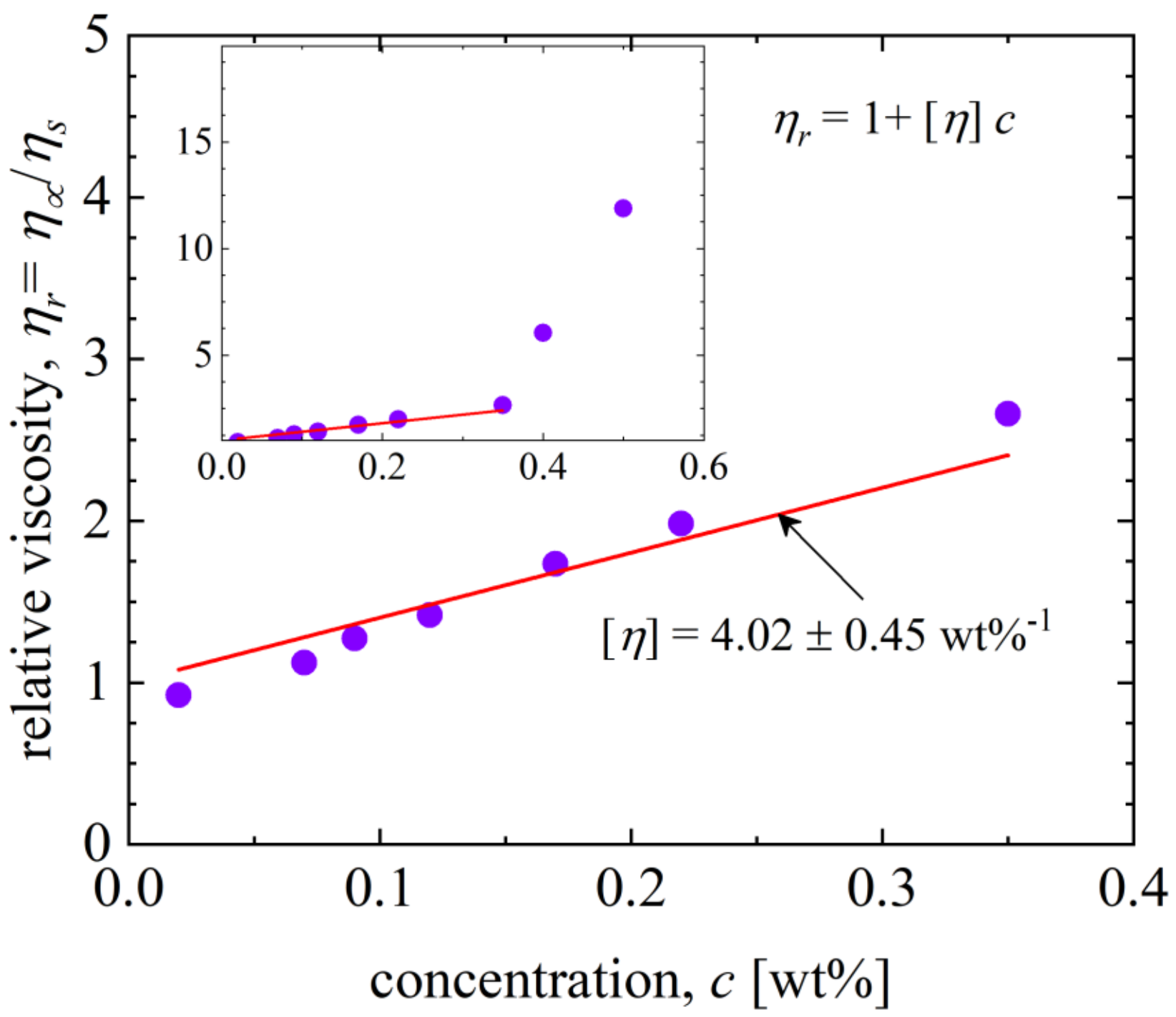

**Figure S2** - At low concentrations, the relative viscosity $\eta_r = \eta_\infty/\eta_s$ at infinite shear rate (obtained using a Carreau-Yasuda model fits, $\eta(\dot{\gamma}) = \eta_\infty + (\eta_0 - \eta_\infty)[1 + (k\dot{\gamma})^a]^{\frac{n-1}{a}}$) agrees well with the Einstein equation ($\frac{\eta}{\eta_s} = 1 + 2.5\phi$ ). For dilute suspensions ($c \to 0$ ), the effective volume fraction can be related to the mass fraction using, $2.5\phi = [\eta]c$, where [η] is the intrinsic viscosity ($[\eta] = 4.02 \pm 0{:}45\; wt\%^{-1}$). The solvent viscosity, $\eta_s$, is taken as that of deionized water (=0.001 $Pa.s$). At higher concentrations ($c > 0.35\; wt\%$) the viscosity strongly deviates in an upward direction due to inter-particle repulsions, consistent with our observation of a measurable linear elastic moduli at $c = 0.4\; wt\%$.

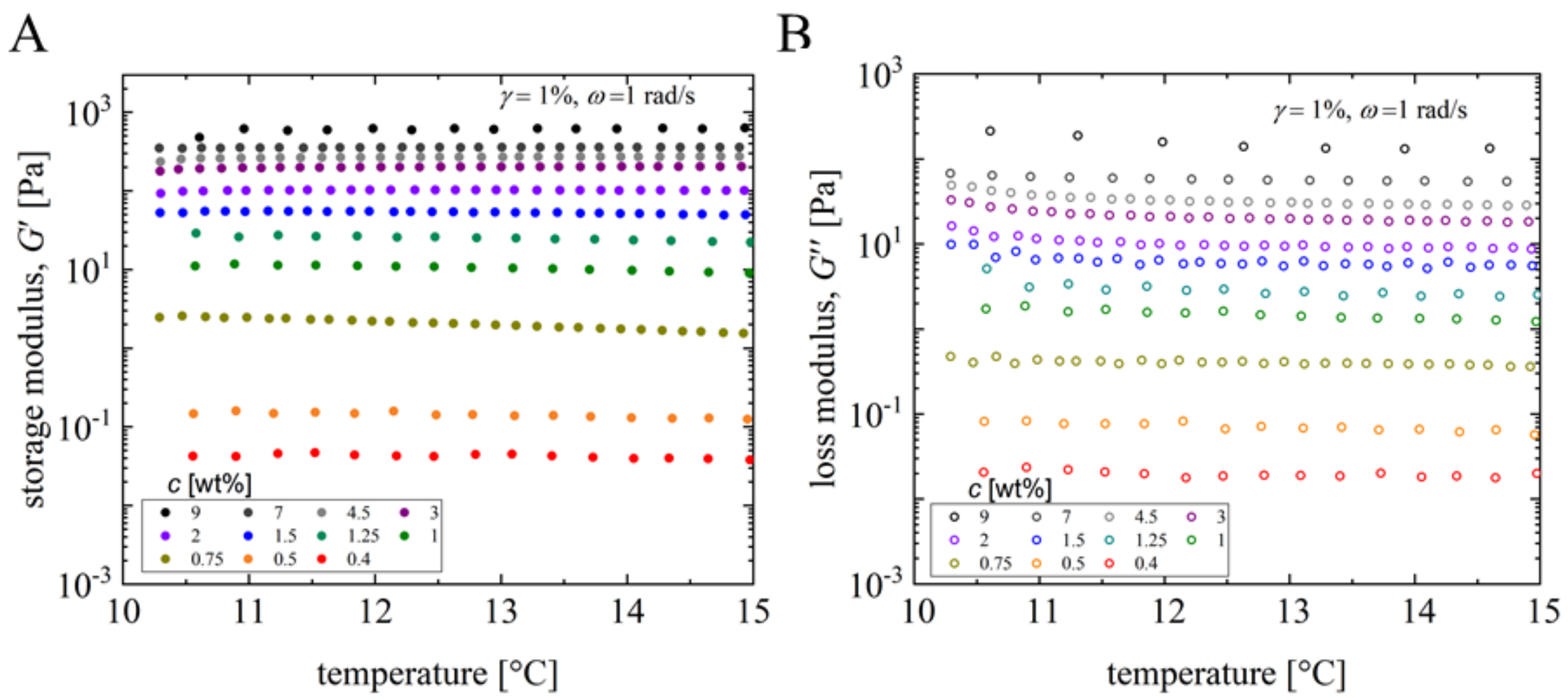

**Figure S3** - (A) Storage modulus, $G'$ and (B) Loss modulus, $G''$, for various microgel concentrations in the temperature range $(10-15)$℃ probed at a fixed strain amplitude of $\gamma_0 = 1\%$ in the linear response regime at an angular frequency of $\omega = 1\, rad/s$ The temperature is increased at a rate of $1$ ℃/min. The rheological properties are temperature independent in the range of probed temperature.

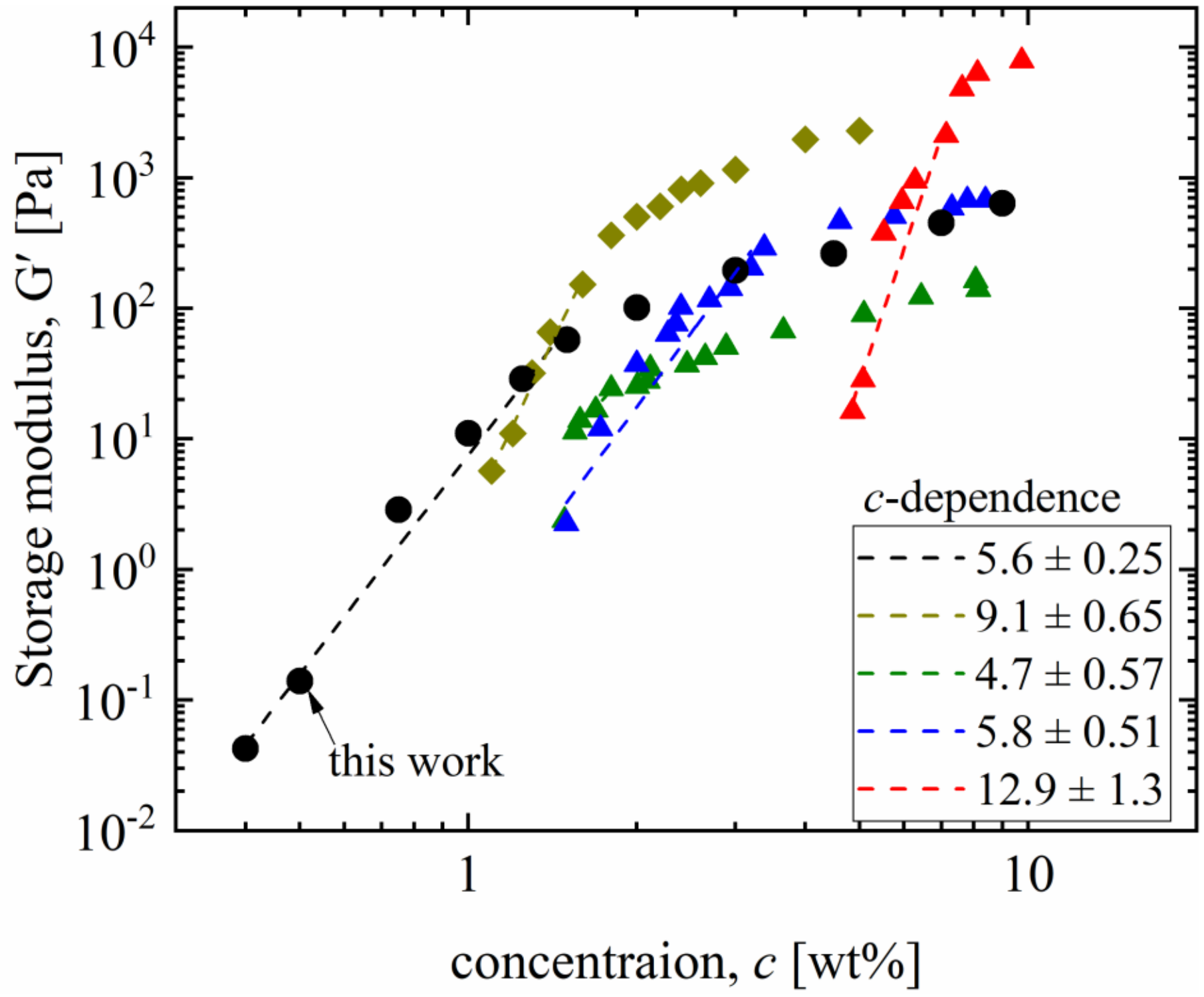

**Figure S4 -** Comparison of the concentration dependent storage modulus as observed in the current work that employs self crosslinked neutral microgel suspensions (black circles) and prior studies of cross-linked ionic microgels (yellow diamonds [11] and blue, green and red triangles [5]). A wide concentration range spanning the glassy and "soft jammed" regimes is shown for all the data with different concentration dependences of shear modulus in the glassy regime. A qualitative universality exists for soft microgels in the sense that, independent of chemistry, all soft particles show a stronger concentration dependence in the glassy regime and roughly linear growth in the "soft jammed" regime. However, the apparent power laws and soft jamming crossover points are highly variable, depending on microgel chemistry, preparation protocol, their internal crosslink density, and the nature of the steric and/or ionic driven deswelling behavior.

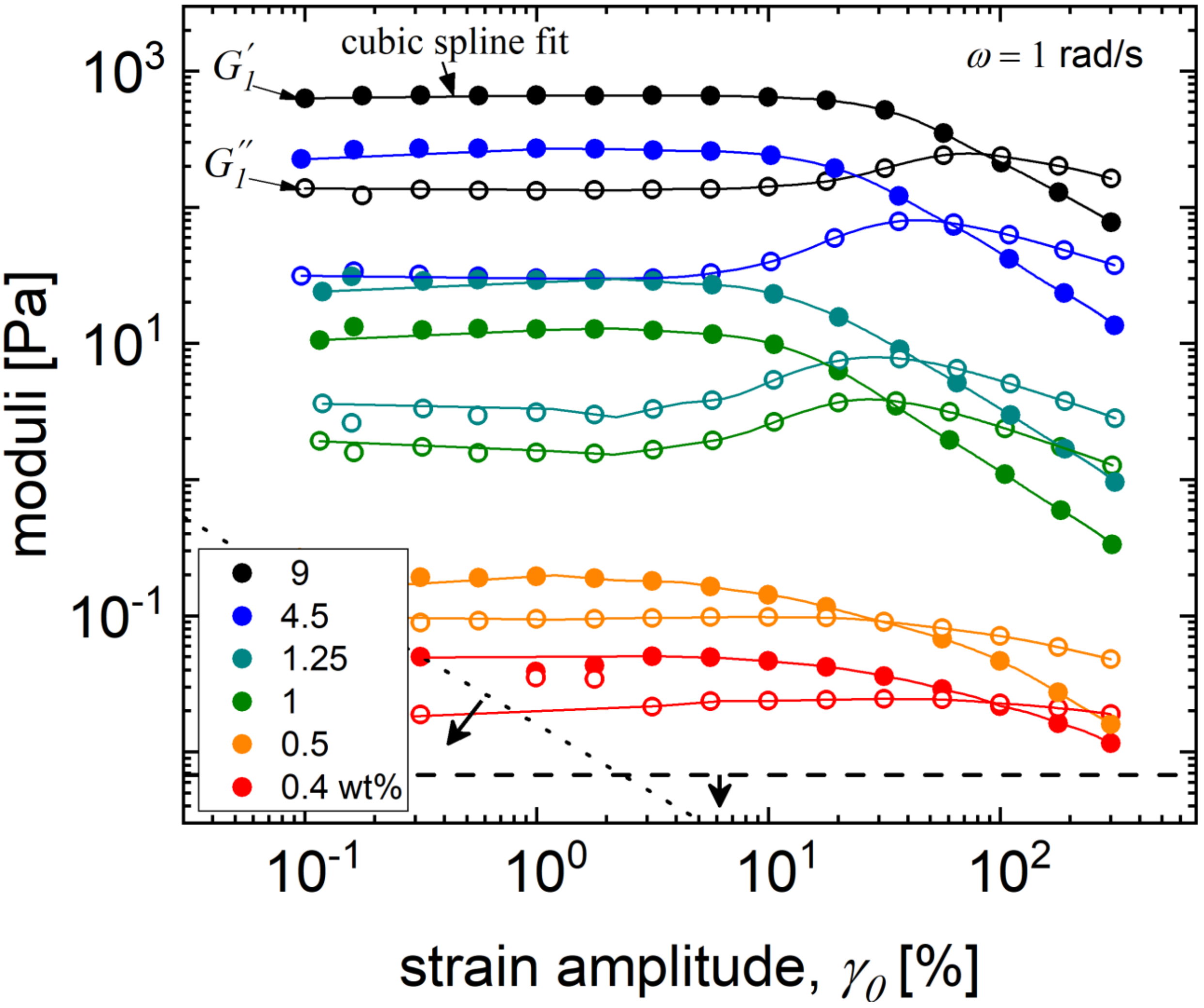

**Figure S5 -** Cubic spline fits to the amplitude sweep data to extract the yield properties. The strain amplitude at which a cubic spline fit to $G_1''$ achieves a maximum is taken as the dynamic yield strain and the point of intersection of cubic spline fits to $G_1'$ *and* $G_1''$is taken as the absolute yield strain.